# Experimental Investigation of Proppant Flow and Transport Dynamics Through Fracture Intersections

Wenpei Ma[1,*], Justin Perng[1] and Ingrid Tomac[1]

**Abstract**

This paper investigates proppant flow and transport in intersected fractures at angles typical for intersections of pre-existing and new hydraulic fractures. Proppant is small granular material, which is placed into hydraulic fractures during geothermal and hydrocarbon reservoir stimulation and props the fluid paths open during reservoir exploitation. This study uses plexiglas laboratory slot experiments enhanced with an advanced image analysis for identifying particle trajectories and quantifying slurry velocities. Although proppant flow and transport has been broadly studied, the effects of intersecting fracture angles have not, especially coupled with fluid viscosities, flow rates, and proppant volumetric concentration effects. This paper specifically investigates the role of intermediate fracture angles, which have been identified to occur most frequently when the new hydraulic fractures intercept the existing ones. Results show that proppant flow and transport behavior after the intersection is very sensitive to carrying fluid viscosity and flow rates alteration, while differentiating proppant volumetric concentrations have a limited effect. Fracture intersection angle itself has a clear effect on proppant flow velocities and proppant settlement; furthermore, it enhances the effects from fluid viscosity, fluid flow rates, and proppant volumetric concentrations.

[1] Structural Engineering Department, University of California San Diego, 9500 Gilman Drive, La Jolla, CA, 92093
* Corresponding Author; Email Address: w6ma@eng.ucsd.edu; Address: Structural Engineering Department, University of California San Diego, 9500 Gilman Drive, La Jolla, CA, 92093

**1.0 Introduction**

This paper evaluates the efficiency of proppant flow and transport at the intersection of a pre-existing fracture and a newly formed hydraulic fracture. Proppant is small, synthetic, or natural, granular material widely used in gas and oil industry, and geothermal reservoirs during hydraulic fracturing for permeability enhancement and production improvement of reservoirs. Proppant is pumped together with fracturing fluid and subsequently placed into fractures to keep them open for a long-term reservoir exploitation. Although many researchers have investigated flow and transport of proppant into planar and simplified fractures, the flow and transport of proppants in different shapes of fracture systems has not yet been fully understood. Simple planar newly formed fractures could easily evolve to complex system while interacting with existing fractures, which has been shown in recent numerical and experimental efforts (Zhang et al., 2007; Dayan et al. 2009; Sahai, 2012; Wong et al., 2013; Sahai et al., 2014; Aimene & Ouenes, 2015; Alotaibi & Miskimins, 2015; Li et al., 2016; Luo & Tomac, 2018a; Luo & Tomac, 2018b; Tong & Mohanty, 2016; Wen et al., 2016; Zou et al., 2016; Chang et al., 2017; Kesireddy, 2017; Kou et al., 2018; Pan et al., 2018; Fjaestad & Tomac, 2019; Kumar & Ghassemi, 2019; Sahai & Moghanloo, 2019; Hampton et al., 2019; Nandlal & Weijermars, 2019).

Laboratory experiments of proppant flow and transport in planar hydraulic fractures have been performed for several decades. Researchers have identified and developed many major observations, conclusions, and theories about proppant settling, flow and transport. However, fractures in reality cannot be as planar, linear and smooth as ideal. An experiment on complex fractures was first conducted by Dayan et al. (2009). A few more studies performed parametric investigated factors afterwards on parametric experiments that may cause different flow and transport behaviors of proppants (Sahai, 2012; Sahai et al., 2014; Alotaibi & Miskimins, 2015; Li et al., 2016; Tong & Mohanty, 2016; Wen et al., 2016; ; Kesireddy, 2017; Pan et al., 2018; Sahai & Moghanloo, 2019). Dayan et al (2009) found out that proppant will not flow into secondary fractures if the flow rate in the primary fracture is too low and until proppant settled enough in the primary fractures. Sahai (2012) observed proppant falling into secondary fracture from primary fracture to be purely due to gravity effects; in addition, the proppant concentration has small effect on the secondary fractures sandbank height compared with flow rate, smaller proppants segregate more significantly and are easier to transport into secondary fractures and secondary fractures that are closer to the wellbore injection point will have more proppants transported into. Sahai et al. (2014) found out that proppant travel efficiency through secondary fractures is related to fluid flow rate, proppant concentration, and proppant size. Li et al. (2016) studied the effect of angles in Y-like shaped intersection, where two of the fractures in the same plane are called primary fractures. The dune height in secondary fracture therefore decreases, and the total propped area along primary fractures increases as the intersection angle between primary and secondary fractures increases from 30° to 90°. For 30° some proppants continue moving in the original flow direction, some of the proppant flows into the intersected fracture. For 90° most proppants continue moving in the original flow direction, only a few flows into the secondary fracture. In that way, dune height in the secondary fracture decreases as intersection angle increases. Tong & Mohanty (2016) confirmed the results done by Li et al. (2016) for the intersection angle beyond 90°. Wen et al. (2016) found out that there is an immediate sandbank height change right after the 90° intersection corner, which is more significant if the intersection is closer to wellbore injection location.

Viscosity is identified as a dominant parameter for proppant transport in a secondary fracture closer to the injection location, while gravity plays a more important role for secondary fractures further away (Wen et al., 2016). Alotaibi & Miskimins (2015) further extended the experiment made by Sahai and Moghanloo before by considering the effect of particle surface roughness and concluded that angular sands have better transport characteristics. Pan et al. (2018) confirmed that fluid flow rate dominates proppant transport in secondary fractures and the proppant settlement length in secondary fractures is inversely proportional to the intersection angle. Kesireddy (2019) also showed how the fluid flow rate dominantly controls the proppant transport in secondary fractures and found that the sandbank height will significantly increase in secondary fractures with increasing primary fracture width.

Although several previous studies investigated the deposited sand dune geometry in complex fracture systems in experimental slots, a relationship between the governing parameters which affect proppant placement efficiency through intersecting fractures at different angles have not yet been quantified. The governing parameters, which have been previously identified to affect the efficiency of proppant flow and transport in complex fracture systems, are intersecting fracture angles, fluid dynamic viscosity, slurry flow rate, and proppant concentration. This paper uses micromechanics and quantifies slurry velocity field using Particle Image Velocimetry (GeoPIV method), which can help understanding of how and to which extent relevant parameters govern the proppant flow and transport through fracture intersections. This study quantifies experimental results at a small scale of detail for two intersecting fracture angles, 30° and 45°.

## 2.0 Methodology

This research uses experimental setup to investigate proppant flow and transport through two plexiglass fractures which contain two different intersecting angles, varying the fluid carrying dynamic viscosity, pumping flow rates and particle volumetric concentrations in slurry. The intersecting angles are chosen as representative for the most common fracture intersections documented in hydraulic fracturing of granite (Frash et al., 2019; Pan et al., 2018, and Li et al., 2016). Proppant particles are injected into a fracture filled with slickwater solution at different flow rates and particle volumetric concentrations. Experiments were recorded with high resolution scientific video cameras, and the conclusions are drawn from visual observations, dune measurements and the GeoPIV video analysis.

### 2.1 Material Selections for Sands and Fluids

Since the focus of this work is to investigate effects of fracture geometry configuration, only one type of sand was selected for all tests, Ottawa F65 sand at varied concentrations. Figure 1 shows the physical appearance of test sands. The sand is round, light-colored, and fine graded with 60/100 meshes, which is a mesh size widely used in hydraulic fracturing.

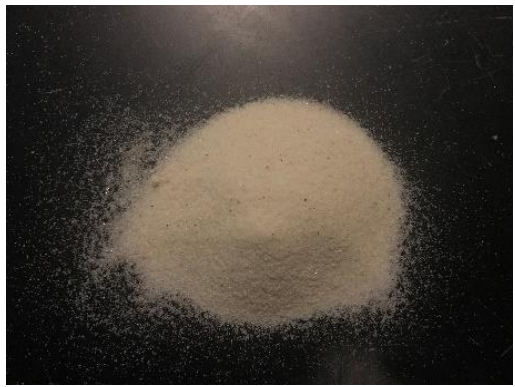

Figure 1. Physical appearance of sand used in the experiment

Besides the major investigation factor of fracture geometry, fluid dynamic viscosity is also varied as it is considered as one of the important investigation factors. The following fluid viscosities are used: water with 0.001 Pa·s (1.0 cp), and glycerol-water mixes with 0.005 Pa·s (5.0 cp) and 0.01 Pa·s (10.0 cp). To make 0.005 Pa·s and 0.01 Pa·s Newtonian fluid, water is slowly mixed with glycerin. The dynamic viscosity of water-glycerin mixture is calculated by Cheng's Method (2008) using Eqns. 1 to 4, and further measured and verified in a rheometer before experiments. Equations 1 to 4 below describe the final dynamic viscosity of the mixture:

$$\mu = \mu_w^\alpha \cdot \mu_g^{1-\alpha} \quad (1)$$

$$\alpha = 1 - C_m + \frac{a \cdot b \cdot C_m \cdot (1 - C_m)}{a \cdot C_m + b \cdot (1 - C_m)} \quad (2)$$

$$a = 0.705 - 0.0017 \cdot T \quad (3)$$

$$b = (4.9 + 0.036 \cdot T) \cdot a^{2.5} \quad (4)$$

where, $\mu$ is the dynamic viscosity of the mixture, $\mu_w$ is the dynamic viscostiy of water, $\mu_g$ is the dynamic viscostiy of glycerin, $\alpha$ is a weighting factor which is a function of glycerin mass concentration ($C_m$), $T$ is temperature and other emprical coefficients ($a$, $b$). To get the empirical coefficients, Cheng (2008) further referred experimental results collected by Segur and Oberstar (1951), as shown in equations (3) and (4) above. Both *a* and *b* depend on temperature of the mixture, here assumed to be the room temperature. A total of 21 L of mixture are required to perform a single test. Once total volume and overall mixture dynamic viscosity are known, volumes of water and pure glycerin could be back-calculated according to the the equations shown above. 11.7 L of water and 9.3 L of glycerin were used to make 0.005 Pa·s viscouse fluid. The density of final mixture is 1130 kg/m³. 9.2 L of water and 11.8 L of glycerin were used to make 0.01 Pa·s viscous fluid and the corresponding density of final mixture is 1161 kg/m³.

Particle Reynolds Number is an important indicator showing how particles move in a suspension during the slurry transport and sediment process. It is a function of the particle diameter, the fluid density, the particle horizontal velocity and the fluid viscosity. Particle Reynolds number is defined in following equation:

$$Re = \frac{\rho_f \cdot v_h \cdot d_s}{\mu_f} \quad (5)$$

where $\rho_f$ is the fluid density, $v_h$ is the particle horizontal velocity, $d_s$ is the particle diameter, and $\mu_f$ is the fluid dynamic viscosity. According to Equation 5, an average particle Reynold numbers for all experiments are calculated and shown as in Table 1 below.

Table 1 Particle Reynolds numbers for all experiments, where *SD* denotes the standard deviation

| EXP | Fluid Density | Mean Particle Diameter | Fluid Viscosity | Mean Particle Velocity | Velocity (*SD*) | Particle Reynolds Number | Reynolds Number (*SD*) |
|---|---|---|---|---|---|---|---|
| | [g/cm³] | [cm] | [g/cm·s] | [cm/s] | [cm/s] | [-] | [-] |
| 01 | 1.00 | 0.02 | 0.01 | 7.24 | 2.00 | 14.47 | 4.00 |
| 02 | 1.00 | 0.02 | 0.01 | 5.26 | 1.18 | 10.52 | 2.37 |
| 03 | 1.00 | 0.02 | 0.01 | 4.77 | 2.46 | 9.53 | 4.92 |
| 04 | 1.00 | 0.02 | 0.01 | 4.55 | 2.49 | 9.09 | 4.98 |
| 05 | 1.00 | 0.02 | 0.01 | 7.98 | 3.54 | 15.96 | 7.09 |
| 06 | 1.00 | 0.02 | 0.01 | 5.97 | 2.79 | 11.93 | 5.59 |
| 07 | 1.00 | 0.02 | 0.01 | 9.40 | 2.20 | 18.80 | 4.39 |
| 08 | 1.00 | 0.02 | 0.01 | 8.80 | 3.25 | 17.61 | 6.50 |
| 09 | 1.13 | 0.02 | 0.05 | 1.84 | 0.58 | 0.83 | 0.26 |
| 10 | 1.13 | 0.02 | 0.05 | 1.38 | 0.72 | 0.63 | 0.32 |
| 11 | 1.13 | 0.02 | 0.05 | 2.52 | 0.81 | 1.14 | 0.37 |
| 12 | 1.13 | 0.02 | 0.05 | 1.55 | 0.60 | 0.70 | 0.27 |
| 13 | 1.13 | 0.02 | 0.05 | 1.10 | 0.47 | 0.50 | 0.21 |
| 14 | 1.13 | 0.02 | 0.05 | 1.33 | 1.14 | 0.60 | 0.51 |
| 15 | 1.13 | 0.02 | 0.05 | 1.91 | 1.19 | 0.86 | 0.54 |
| 16 | 1.13 | 0.02 | 0.05 | 1.46 | 1.27 | 0.66 | 0.57 |
| 17 | 1.16 | 0.02 | 0.10 | 1.89 | 0.30 | 0.44 | 0.07 |
| 18 | 1.16 | 0.02 | 0.10 | 1.50 | 0.33 | 0.35 | 0.08 |
| 19 | 1.16 | 0.02 | 0.10 | 2.64 | 0.30 | 0.61 | 0.07 |
| 20 | 1.16 | 0.02 | 0.10 | 1.89 | 0.28 | 0.44 | 0.06 |
| 21 | 1.16 | 0.02 | 0.10 | 1.80 | 0.26 | 0.42 | 0.06 |
| 22 | 1.16 | 0.02 | 0.10 | 1.35 | 0.25 | 0.31 | 0.06 |
| 23 | 1.16 | 0.02 | 0.10 | 2.17 | 0.34 | 0.50 | 0.08 |
| 24 | 1.16 | 0.02 | 0.10 | 1.54 | 0.24 | 0.36 | 0.06 |

Particle Stokes Number is another important indicator on particle settlement behavior during the slurry transport and sediment process. It is a function of the particle density, the particle vertical velocity, the particle diameter, the fracture aperture, and the fluid dynamic viscosity. Particle Stokes number is defined by following equation:

$$St = \frac{\rho_s \cdot v_i \cdot d_s^2}{18 \cdot w \cdot \mu_f} \quad (6)$$

where, $\rho_s$ is the particle density, $v_i$ is the average particle vertical velocity, $d_s$ is the particle diameter, $w$ is the fracture aperture, and $\mu_f$ is the fluid viscosity. According to Equation 6, an average particle Stokes number for all experiments are calculated and shown as in Table 2 below.

Table 2 Particle Stokes numbers for all experiments, where *SD* denotes the standard deviation

| EXP | Fluid Density | Mean Particle Diameter | Particle Density | Fracture Aperture | Fluid Viscosity | Mean Particle Velocity | Velocity (*SD*) | Particle Stokes Number | Stokes Number (*SD*) |
|---|---|---|---|---|---|---|---|---|---|
| | [g/cm³] | [cm] | [g/cm³] | [cm] | [g/cm·s] | [cm/s] | [cm/s] | [-] | [-] |
| 01 | 1.00 | 0.02 | 2.65 | 0.60 | 0.01 | 2.27 | 1.23 | 0.0223 | 0.0120 |

| | | | | | | | | |
|---|---|---|---|---|---|---|---|---|
| 02 | 1.00 | 0.02 | 2.65 | 0.60 | 0.01 | 1.11 | 0.40 | 0.0109 | 0.0039 |
| 03 | 1.00 | 0.02 | 2.65 | 0.60 | 0.01 | 1.39 | 1.09 | 0.0136 | 0.0107 |
| 04 | 1.00 | 0.02 | 2.65 | 0.60 | 0.01 | 1.57 | 0.90 | 0.0155 | 0.0088 |
| 05 | 1.00 | 0.02 | 2.65 | 0.60 | 0.01 | 1.51 | 1.19 | 0.0148 | 0.0117 |
| 06 | 1.00 | 0.02 | 2.65 | 0.60 | 0.01 | 1.32 | 1.04 | 0.0129 | 0.0102 |
| 07 | 1.00 | 0.02 | 2.65 | 0.60 | 0.01 | 2.26 | 0.91 | 0.0222 | 0.0089 |
| 08 | 1.00 | 0.02 | 2.65 | 0.60 | 0.01 | 1.36 | 1.09 | 0.0133 | 0.0107 |
| 09 | 1.13 | 0.02 | 2.65 | 0.60 | 0.05 | 0.20 | 0.25 | 0.0004 | 0.0005 |
| 10 | 1.13 | 0.02 | 2.65 | 0.60 | 0.05 | 0.07 | 0.27 | 0.0001 | 0.0005 |
| 11 | 1.13 | 0.02 | 2.65 | 0.60 | 0.05 | 0.18 | 0.30 | 0.0004 | 0.0006 |
| 12 | 1.13 | 0.02 | 2.65 | 0.60 | 0.05 | 0.10 | 0.29 | 0.0002 | 0.0006 |
| 13 | 1.13 | 0.02 | 2.65 | 0.60 | 0.05 | 0.09 | 0.22 | 0.0002 | 0.0004 |
| 14 | 1.13 | 0.02 | 2.65 | 0.60 | 0.05 | 0.10 | 0.29 | 0.0002 | 0.0006 |
| 15 | 1.13 | 0.02 | 2.65 | 0.60 | 0.05 | 0.02 | 0.32 | 0.0000 | 0.0006 |
| 16 | 1.13 | 0.02 | 2.65 | 0.60 | 0.05 | 0.17 | 0.38 | 0.0003 | 0.0007 |
| 17 | 1.16 | 0.02 | 2.65 | 0.60 | 0.10 | 0.07 | 0.21 | 0.0001 | 0.0002 |
| 18 | 1.16 | 0.02 | 2.65 | 0.60 | 0.10 | 0.09 | 0.19 | 0.0001 | 0.0002 |
| 19 | 1.16 | 0.02 | 2.65 | 0.60 | 0.10 | 0.14 | 0.20 | 0.0001 | 0.0002 |
| 20 | 1.16 | 0.02 | 2.65 | 0.60 | 0.10 | 0.13 | 0.22 | 0.0001 | 0.0002 |
| 21 | 1.16 | 0.02 | 2.65 | 0.60 | 0.10 | 0.18 | 0.19 | 0.0002 | 0.0002 |
| 22 | 1.16 | 0.02 | 2.65 | 0.60 | 0.10 | 0.12 | 0.20 | 0.0001 | 0.0002 |
| 23 | 1.16 | 0.02 | 2.65 | 0.60 | 0.10 | 0.06 | 0.23 | 0.0001 | 0.0002 |
| 24 | 1.16 | 0.02 | 2.65 | 0.60 | 0.10 | 0.16 | 0.24 | 0.0002 | 0.0002 |

## 2.2 Experimental Setup

Two fractures are designed for the experiment. Figures 2a and 2b show the top views of fractures with intersecting angles at 30° and 45°, inspired by observations from Frash et al. (2019), Pan et al. (2018), and Li et al. (2016). The entire fracture system includes five parts: the entrance funnel, the entrance fracture, the middle fracture, the exit fracture, and the exit funnel. The intersection angle is defined as the angle between the direction of fluid flow in entrance/exit fracture and the direction of fluid flow in middle fracture. The exit and entrance funnels have a slope of 5°. The entrance, middle, and exit fracture are 203 mm high and 6 mm wide.

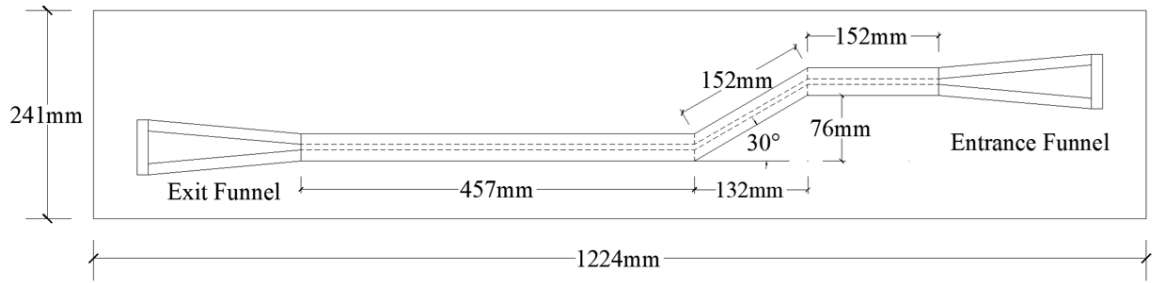

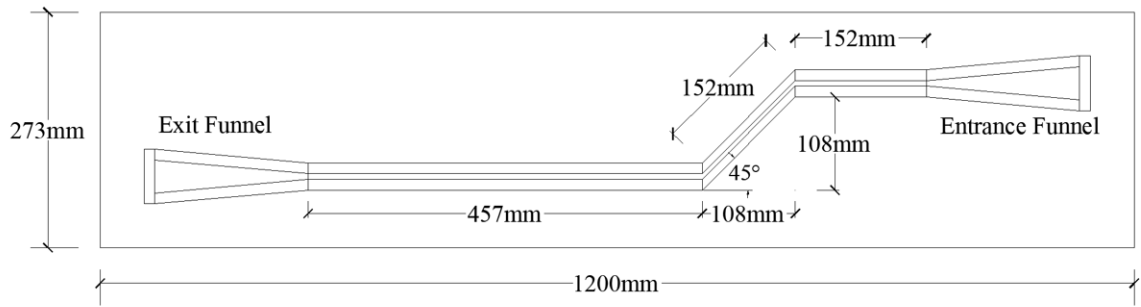

Figure 2. Plexiglass fractures top view, a) 30° intersecting angle and b) 45° intersecting angle

    Figure 3 shows the physical appearance of the 30° designed fracture. Most of the acrylic plates are completely glued with each other, except the removable top covers of entrance and exit funnels. Rubber bands are used for sealing. There are three holes on each side of the fracture. Fluids were injected into the middle level and expelled from the upper level.

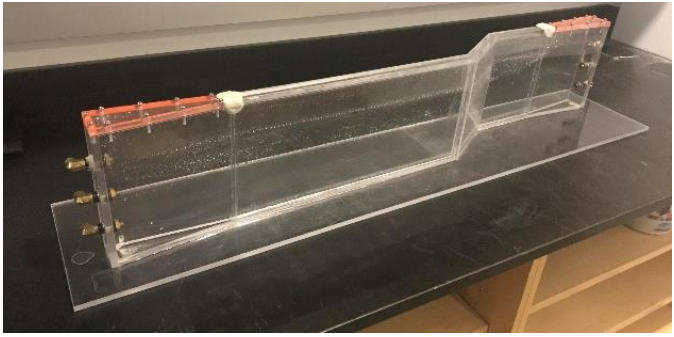

Figure 3 Physical appearance of the designed fracture

Figure 4 shows the configuration of the entire experiment system. The sand and fluid are continuously mixed in a bucket at the very top using electrical mixer at sufficiently high rate to keep proppant particles suspended in fluid. Then, the slurry is pumped and injected into the fracture from the mid-level hole of the entrance funnel. All exiting fluids are collected into the bucket on the table from the top-level hole of the exit funnel. To better record the sand flow in the fracture, black background paper is put on the back of the fracture, from the entrance part to middle and exit part, not including the funnels. Cameras are placed in front of the fracture. To get the best video quality, the distance between the camera lenses and the fracture face is generally around 140 cm.

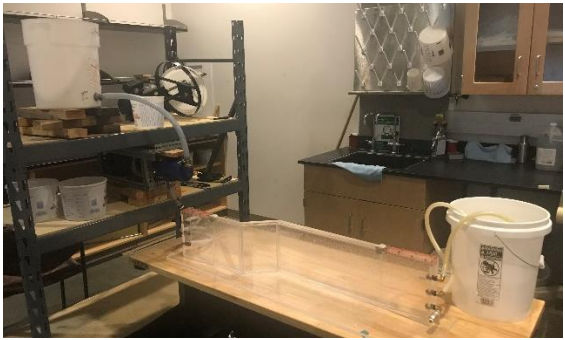

Figure 4 Physical configuration of the experiment system

Three cameras are used to record each test. A SONY DSC-RX10M3 digital still camera is used to record particle flow process in entrance funnel and entrance fracture part of the fracture system. Videos were shot at 1920x1080 pixels 60 fps. A Nikon D160 digital camera was used to record particle flow process in entire exit fracture part of the fracture system. Videos were shot at 1280x720 pixels 60 fps. A third high-speed Phantom C320 camera was used to record first half of the exit fracture part and further analyzed by GeoPIV Method. All videos were shot 1280x1024 pixels 900 fps.

Beside the main investigation parameter of fracture intersection angle, the experiments also consider the following factors: fluid viscosity, volumetric concentration of sand, and fluid flow rate. Table 3 below shows all tests conducted in terms of combinations of different factors. The fluid flow rate is a function of the pump rotor frequency, the fluid viscosity and the volumetric concentration of sand, obtained during pump calibrations. In this experiment, the fluid flow rate is primarily controlled by the pump rotor frequency. Table 4 below shows the relationship between the fluid flow rate and the pump rotor frequency for each case.

Table 3 Experimental cases

| EXP | Particle Type | Fluid Viscosity (Pa·s) | Volumetric Concentration of Sand (%) | Pump Rotor Frequency (Hz) | Fracture Intersection Angle (°) |
|---|---|---|---|---|---|
| 1 | Fine Ottawa Sand | 0.001 | 10 | 10 | 30 |
| 2 | Fine Ottawa Sand | 0.001 | 10 | 10 | 45 |
| 3 | Fine Ottawa Sand | 0.001 | 10 | 20 | 30 |
| 4 | Fine Ottawa Sand | 0.001 | 10 | 20 | 45 |
| 5 | Fine Ottawa Sand | 0.001 | 20 | 10 | 30 |
| 6 | Fine Ottawa Sand | 0.001 | 20 | 10 | 45 |
| 7 | Fine Ottawa Sand | 0.001 | 20 | 20 | 30 |
| 8 | Fine Ottawa Sand | 0.001 | 20 | 20 | 45 |
| 9 | Fine Ottawa Sand | 0.005 | 10 | 10 | 30 |
| 10 | Fine Ottawa Sand | 0.005 | 10 | 10 | 45 |
| 11 | Fine Ottawa Sand | 0.005 | 10 | 20 | 30 |
| 12 | Fine Ottawa Sand | 0.005 | 10 | 20 | 45 |
| 13 | Fine Ottawa Sand | 0.005 | 20 | 10 | 30 |
| 14 | Fine Ottawa Sand | 0.005 | 20 | 10 | 45 |
| 15 | Fine Ottawa Sand | 0.005 | 20 | 20 | 30 |
| 16 | Fine Ottawa Sand | 0.005 | 20 | 20 | 45 |
| 17 | Fine Ottawa Sand | 0.010 | 10 | 10 | 30 |
| 18 | Fine Ottawa Sand | 0.010 | 10 | 10 | 45 |
| 19 | Fine Ottawa Sand | 0.010 | 10 | 20 | 30 |
| 20 | Fine Ottawa Sand | 0.010 | 10 | 20 | 45 |
| 21 | Fine Ottawa Sand | 0.010 | 20 | 10 | 30 |
| 22 | Fine Ottawa Sand | 0.010 | 20 | 10 | 45 |
| 23 | Fine Ottawa Sand | 0.010 | 20 | 20 | 30 |
| 24 | Fine Ottawa Sand | 0.010 | 20 | 20 | 45 |

Table 4 Pump fluid flow rates

| Viscosity (Pa·s) | Volumetric Concentration (%) | Frequency (Hz) | Flow Rate (L/min) |
|---|---|---|---|
| 0.001 | 10 | 10 | 3.90 |
| 0.001 | 10 | 20 | 4.55 |
| 0.001 | 20 | 10 | 3.30 |
| 0.001 | 20 | 20 | 3.70 |
| 0.005 | 10 | 10 | 3.35 |
| 0.005 | 10 | 20 | 3.68 |
| 0.005 | 20 | 10 | 2.55 |
| 0.005 | 20 | 20 | 3.50 |
| 0.010 | 10 | 10 | 2.50 |
| 0.010 | 10 | 20 | 3.20 |
| 0.010 | 20 | 10 | 2.65 |
| 0.010 | 20 | 20 | 3.13 |

## 2.3 GeoPIV Analysis

Advanced Particle Image Velocimetry (PIV) method, adopted for studying particulate materials, is used to analyze the videos recorded by Phantom camera. The main software is GeoPIV-RG, a Matlab module developed by Stainer et al. (2015), previously called GeoPIV by Take & White (2002). Figure 5 below describes the GeoPIV-RG flow chart. Before launching the main process code, it is required to select and decide the frames/pictures to be analyzed, to choose the corresponding regions of interest, and to decide the size and spacing of meshes. The main code tracks particle movements among images by comparing the reference and subsequent images at the point in time of interest. The leapfrog method retains the initial image as a reference image after every computation, which is suggested if there is no control point in the experiment. However, if too many wild results occur, GeoPIV-RG uses a sequential scheme, which updates reference images after every computation. GeoPIV-RG generates a displacement vector field and displacement contours for selected region of interest and the output has a unit of pixels per frame. Since the scientific cameras secure a high-precision relationship between the video/image pixel size and actual dimension length, as well as the relationship between time and the frame recording rate, a velocity field is obtained by a conversion from the displacement field.

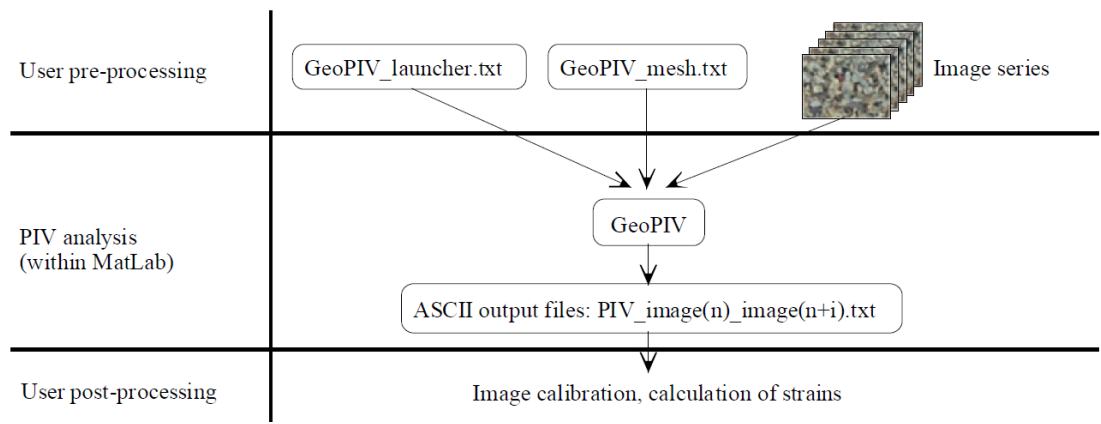

Figure 5 Flow chart of GeoPIV-RG software (adopted from Take & White, 2002)

**2.4 Error Control and Accuracy**

Sources of errors in the GeoPIV-RG analysis can originate from the camera performance, image and video setups, lighting environments, the experiment components setup and the GeoPIV-RG software post-processing. In our experiments, the Phantom camera recorded high-resolution high-frame-rate videos. Two LED lights with 12,000 lumens were placed on both sides of the experiment to enhance lighting conditions and to minimize shadows. To avoid image distortions, the camera lens was set to be in same level and perpendicular to fracture face, and the camera's position was secured with a tripod. Therefore, the GeoPIV-RG analysis remains the major uncertainty source in the experiments. To conclude, a good understanding and use of the GeoPIV-RG analysis will largely ensure the accuracy of results.

The input parameters causing errors of the GeoPIV-RG software depend on the image quality, lighting changes, the image-particle diameter, the spatial variation, recording angles and mesh sizes (White and Take, 2002; Stanier et al., 2015; Luo & Tomac, 2018; Fjaestad & Tomac, 2019). Given experiment conditions stated previously, all hardware-related input parameters, except the mesh sizes, have been controlled by fixing position and ensuring sufficient accuracy. It has been previously found that smaller mesh sizes will provide more local information while larger mesh size will provide better precision (Stanier et al., 2015; Lan, 2016; Fjaestad, 2018). To verify what mesh size will be the best for our experiment analysis, four different sizes were analyzed: 10×10, 20×20, 30×30, and 40×40 pixels. In addition, two neighboring area with similar velocities were also selected for comparison. Mean velocities of each region and mesh size were computed. For both selected regions, the percent difference of results between 10-pixel and 20-pixel mesh are above 10%. However, for both regions, the percent difference for results between 20-pixel and 30-pixel mesh, and 30-pixel and 40-pixel mesh are less than 2%. This indicates that 20-pixel mesh and above will provide more consistent results for the range of experimental conditions in this research. When viewing 10-pixel mesh only, the percent differences between two regions is above 10%. For mesh sizes with 20-pixel and above, the percent differences between two regions is less than 5%. This is another indicator which suggests 20-pixels and above provide more accurate result. It is important to remember that local details were observed during the error analysis.

**3.0 Results**

This section describes the visual inspections of proppant settlements, displacement and velocities from GeoPIV analysis. A correlation between proppant settlement and proppant velocities is investigated considering effects of fracture intersection angles, proppant volumetric concentration, carrying flow rate and dynamic viscosities.

**3.1 Visual observations on effect of carrying dynamic fluid viscosity**

In 0.001 Pa·s fluid, the 45° intersecting fracture has a steeper and more various overall dune shape than a 30° fracture for all of the observed particle concentrations and flow rates in Experiments 01 to 08, as shown in Figures 7a and 7b. For example, the 45° fracture dune angles are measured between 9.6° and 16°; while for 30° fracture, the dune angles are between 6.8° and 9° from horizontal as in Figure 6e. The slope steepness and variety are more clearly observed further down the fracture, in the middle and exit branches. Additionally, the 45° intersecting fracture has a more convex curved slope (see all exit branches in Figure 7a and Figure 7b), which indicates a more rapid settlement right after exiting the intersected fractures, while the 30° intersecting fracture causes a flatter slope. Specifically, 45° intersection fracture causes a small localized 'hump' just at the beginning part of the exit branch for a high volumetric concentration of sands, as shown in circled parts in Figure 8. While maintaining at 10% volumetric proppant concentration in the slurry (experiments 01-04), intersecting angle effect on dune shape angle is slightly less significant than that caused by flow rate. For example, the slope angles are 8.5° and 6.8° for 30° intersecting fractures, 11° and 9.6° for 45° intersecting fractures. The differences in the dune angle are 2.5° and 2.8° for low and high flow rate cases. As sand concentration maintains at 20% (experiments 05-08), the intersection angle starts dominating the shape formation of the dune slope. The differences in dune angles are 4° and 7° for low and high flow rate cases. At both lower flow rate in the slurry (experiments 01, 02, 05, 06) and higher flow rate in the slurry (experiments 03, 04, 07, 08), higher proppant volumetric concentration will help to create a more sloped settlement but not very significantly, especially for the middle branch. As flow rate increases, the significance of the combined effect of the volumetric proppant concentration and the intersection angle ramps up producing increasingly steep dune angles. For lower flow rate conditions, the differences in settlement slopes (30° vs. 45° intersection angle) are 2.5° and 5°. For higher flow rate conditions, the differences in settlement slopes are 2.2° and 7°.

In general, for all 0.005 Pa·s carrying fluid experiments for all flow rates and proppant volumetric concentrations (experiments 09-16), 45° intersecting fractures also have a slightly steeper settlement slope as shown in Figure 9a and 9b, compared with 30° intersecting fractures. For example, the 45° fracture dune angles are measured between 6° and 7.5°; while for 30° fracture, the dune angles are between 5° and 7° from horizontal as in Figure 6f. However, this effect is not as significant as in pure water condition, 0.001 Pa·s carrying fluid (comparing Figure 9a vs. Figure 7a, or Figure 9b vs. Figure 7b), confirming that the increase of carrying fluid dynamic viscosity contributes to better flow and transport. As shown in Figures 9a and 9b, even though the tests have stopped for a while, there are still sands flowing in the fluid. For both 10% (experiments 09-12) and 20% proppant volumetric concentrations (experiments 13-16), the role of flow rate is not as strong as that of intersecting angle, which dominates the formation of settlement slope. At low proppant volumetric concentrations, the dune angle difference is 1° for a lower flow rate and 0° for a higher flow rate. At high proppant volumetric concentrations, the dune angle difference is 2° for a lower flow rate and 0.5° for a higher flow rate. For both low (experiments 09, 10, 13, 14) and high (experiments 11, 12, 15, 16) flow rate conditions, higher proppant volumetric concentrations help to shape a slightly steeper dune, combined with the effect of intersecting angle. Under low flow rate conditions, the dune angle difference is 1° for lower sand ratio and 2° for higher sand ratio. As for the high flow rate cases, the dune angle difference is 0° for lower sand ratio and 0.5° for higher sand ratio. Higher proppant volumetric concentration also helps to create a more various settlement shapes in the entrance branch (see Figures 9a and 9b). The sand dune is flat in the entrance and middle branches for lower sand ratio conditions.

Results under 0.01 Pa·s carrying fluid (experiments 17-24) are alike that of 0.005 Pa·s carrying fluid. General settlement slope shape characterizations are all preserved while considering the effect of sand ratio, flow rate, and most importantly intersecting angles. The only major difference is that the slopes of all 8 cases are further flattened, as shown in Figures 10a and 10b. The 45° fracture dune angles are measured between 1.2° and 3.1°; while for 30° fracture, the dune angles are between 0.3° and 1.3° from horizontal as in Figure 6g.

While ignoring all other factors, an increase in carrying fluid dynamic viscosity progressively flattens the dune in the exit branch, increases the horizontal sand transport, improves proppant transport efficiency, and reduces gravity effects as shown in Figure 6d. The multiphase flow remains conserved from the injection point to the collection point, and sand particles remained floating after pumping stops. If considering effect of intersection angle only as shown in Figure 6a, ignoring all other factors, 45° intersection angle generally provides wider range of proppant settlement slope angle. If considering effect of proppant concentration only (Figure 6b) or fluid flow rate (Figure 6c), those effects has a relatively weak effect on settlement slope comparing with other factors.

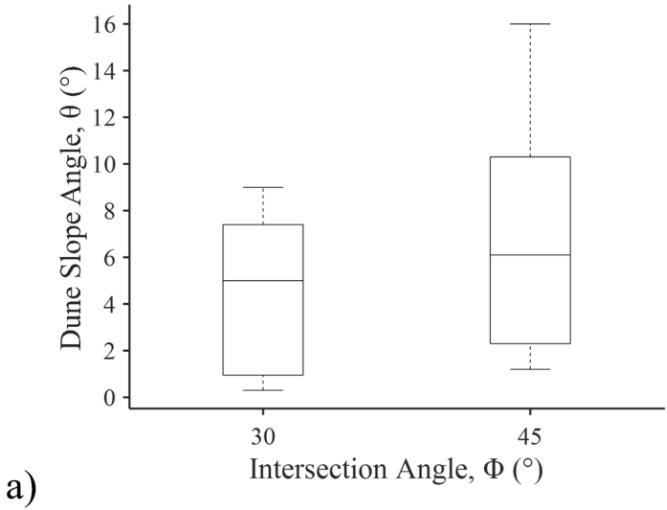

a)

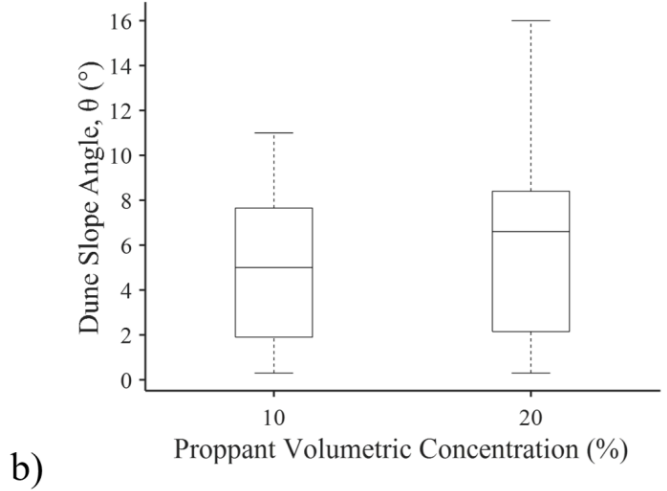

b)

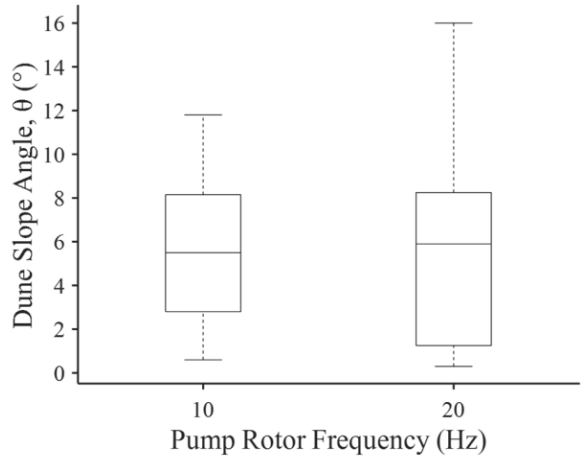

c)

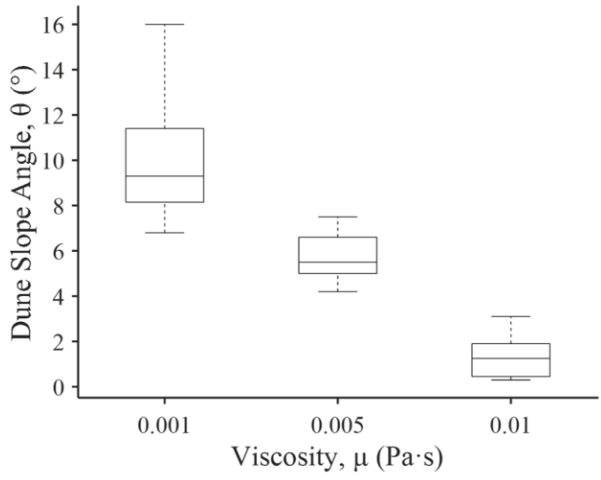

d)

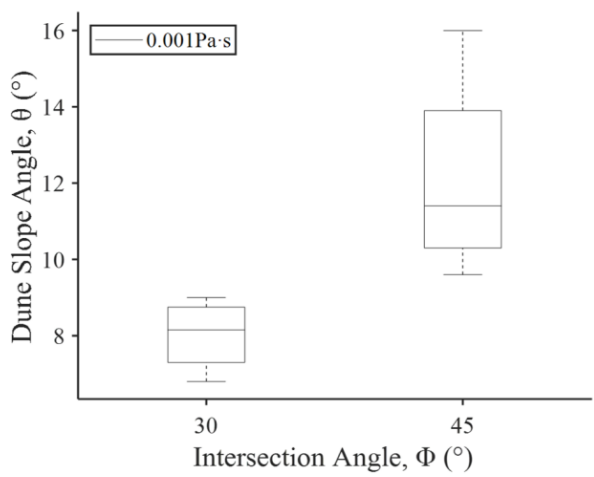

e)

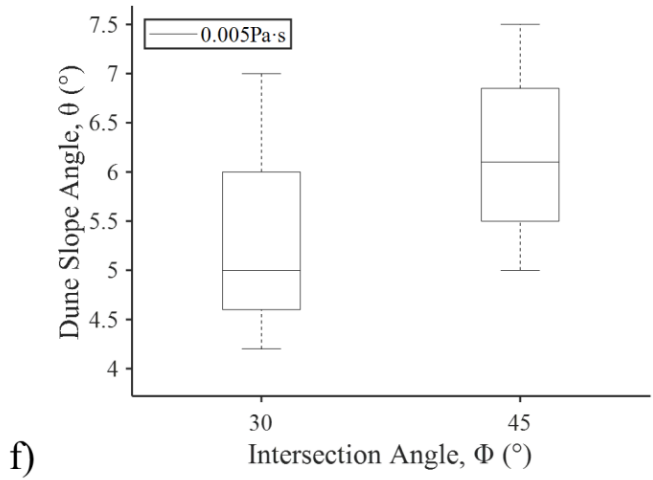

f)

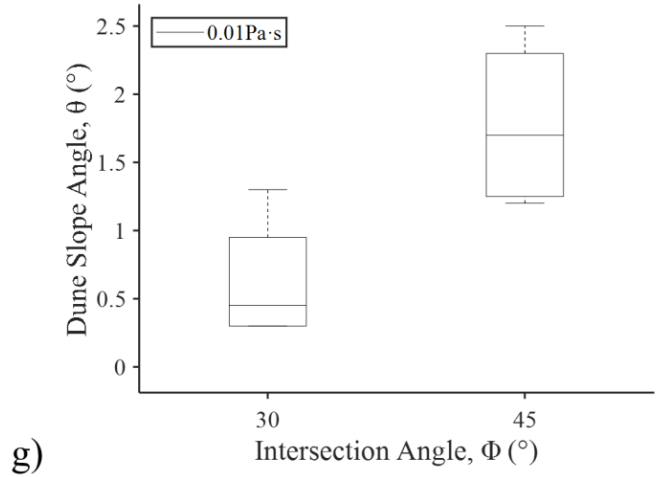

g)

Figure 6 Dune slope angles for all experiments: a) considering effect of intersecting angle only; b) considering effect of proppant concentration only; c) considering effect of fluid flow rate only; d) considering effect of viscosity only; e) under 0.001 Pa·s fluid viscosity; f) under 0.005 Pa·s fluid viscosity; g) under 0.01 Pa·s fluid viscosity

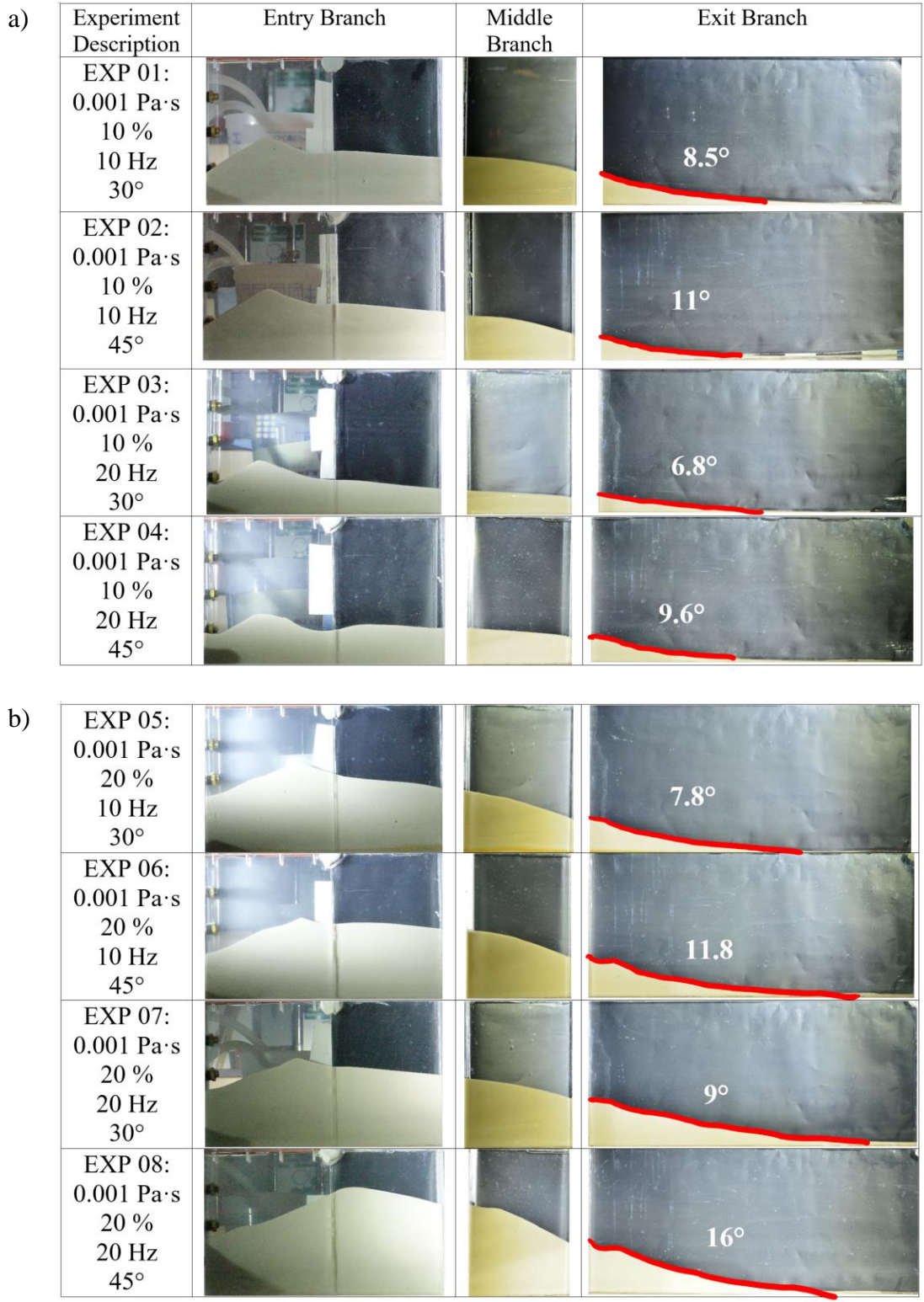

Figure 7 a) Resulting dune settlement shapes under 0.001 Pa·s carrying fluid, b) continued

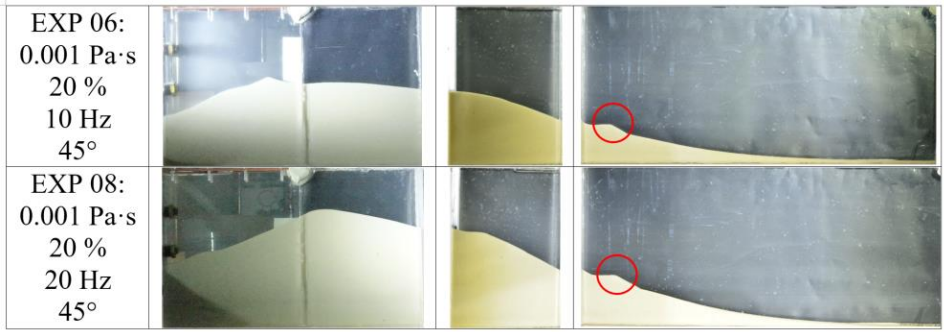

Figure 8) Resulting 'hump' shape under 0.001 Pa·s carrying fluid

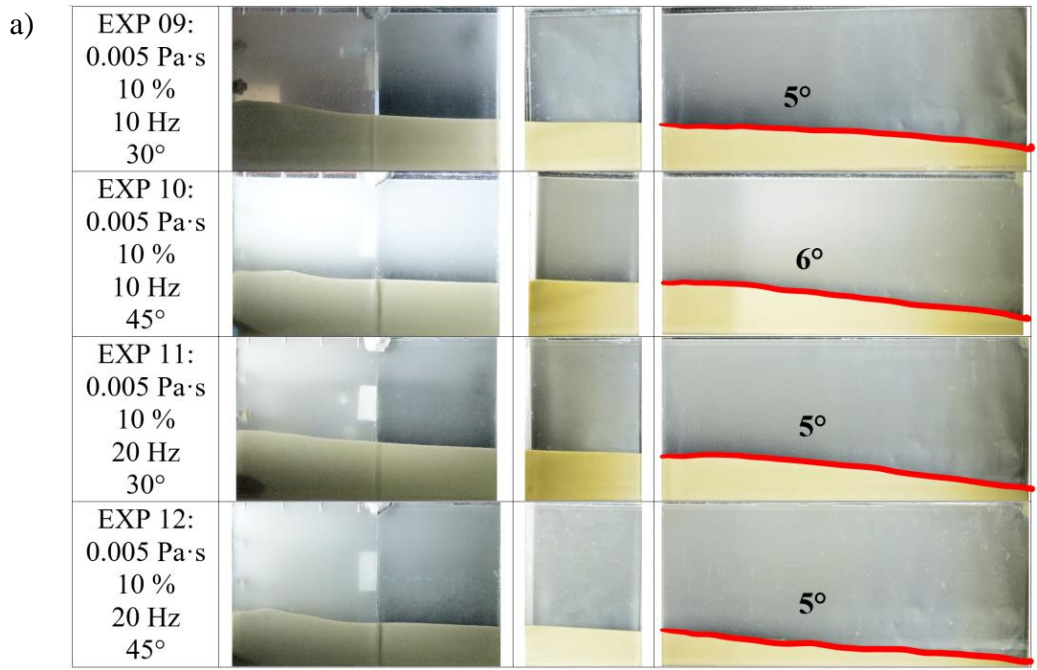

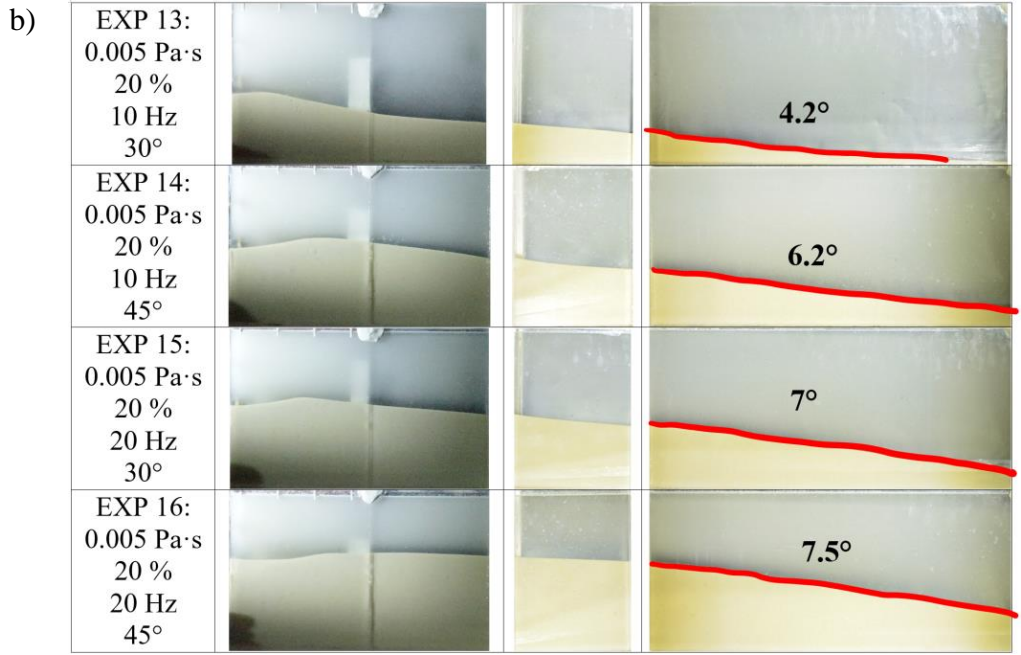

Figure 9 a) Resulting dune settlement shapes under 0.005 Pa·s carrying fluid, b) continued

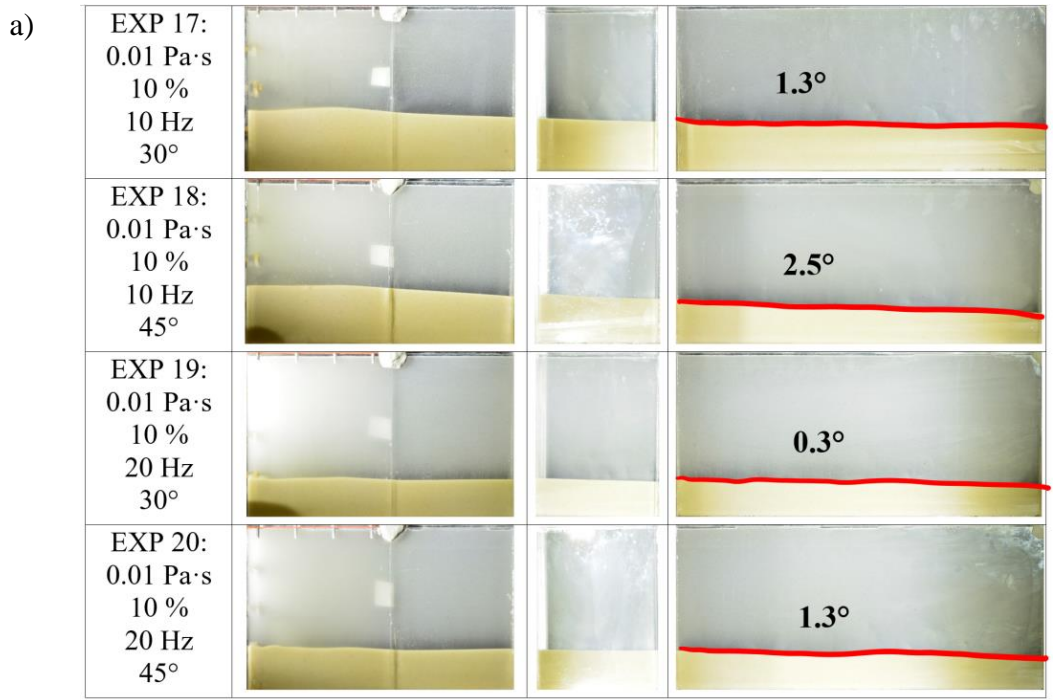

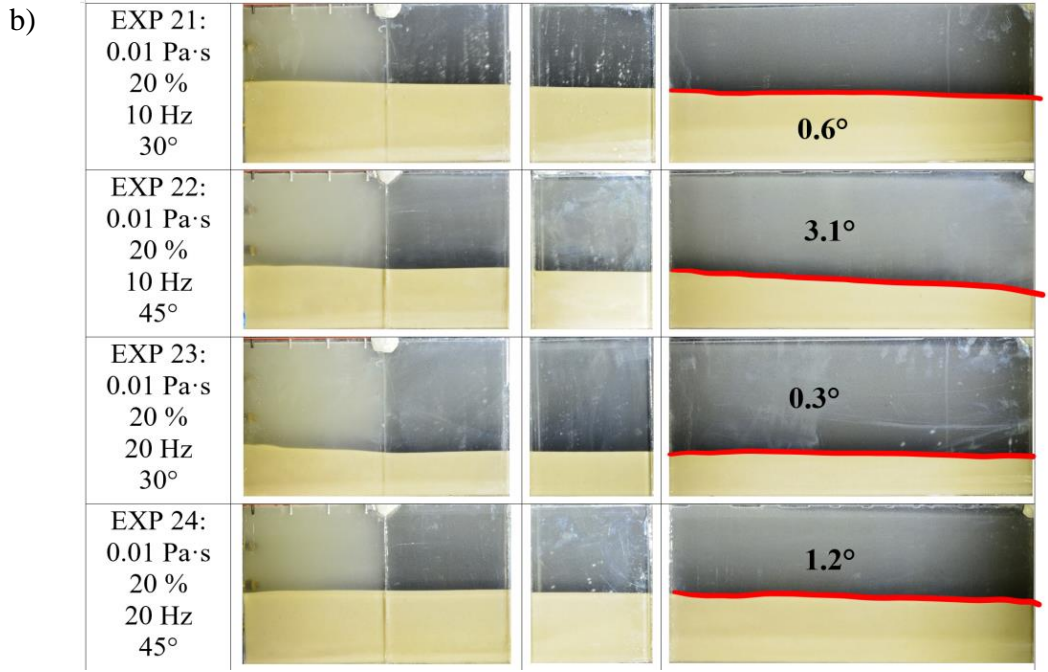

Figure 10 a) Resulting dune settlement shapes under 0.01 Pa·s carrying fluid, b) continued

## 3.2 Proppant Displacement and Velocity Measurements

To quantify the effects of governing parameters for better understanding of proppant flow and transport through intersected fractures, GeoPIV-RG analyses are performed for governing cases. Figures 11a to 11c show selected results (experiments 08, 16, and 24) from GeoPIV-RG. Four types of graphs are included: horizontal displacement contour, vertical displacement contour, resultant displacement contour, and displacement vector field. Color scale in Figure 11a to 11c represents displacement magnitudes, which are here presented in pixels. For the velocity field calculation, image pixels are related to the experiment length dimensions. Displacement vector field is shown to illustrate analysis of one specific rectangular region. Figures 12a and 12b show post-processed information for selected experiments including colored resultant velocity vector field, velocity range boxplot, and velocity vector direction histogram for the same selected experiments.

a)

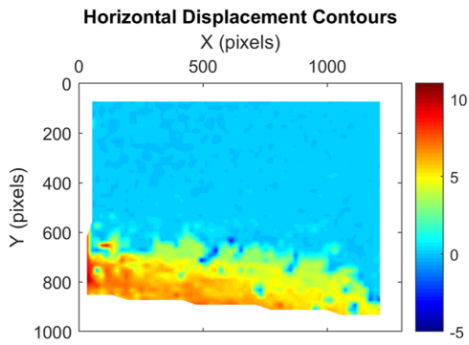 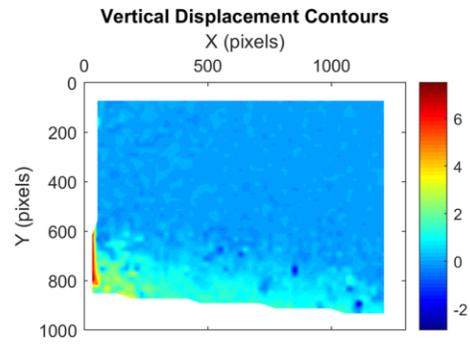

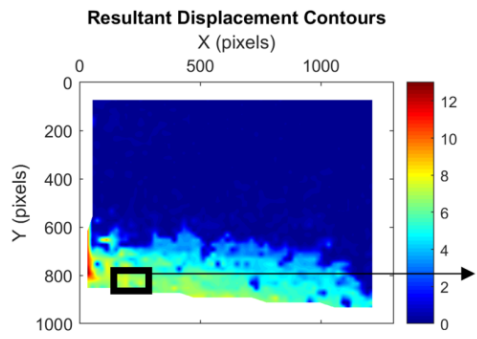 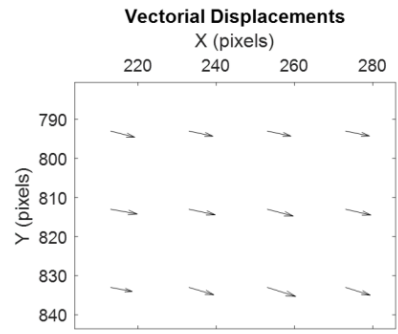

b)

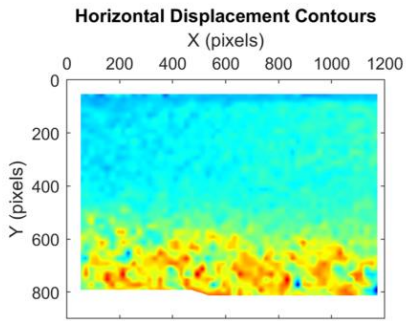 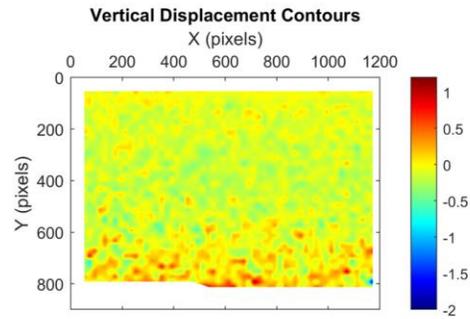

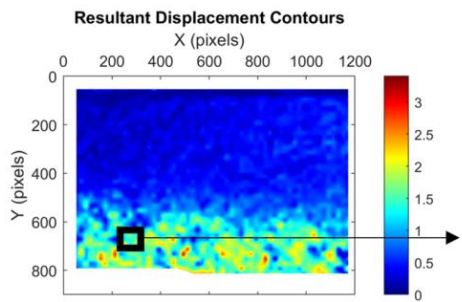 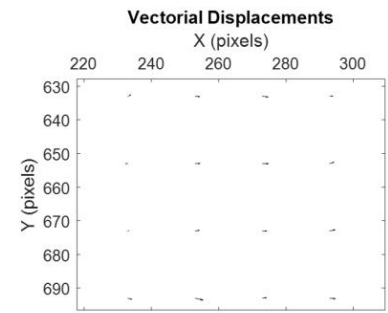

c)

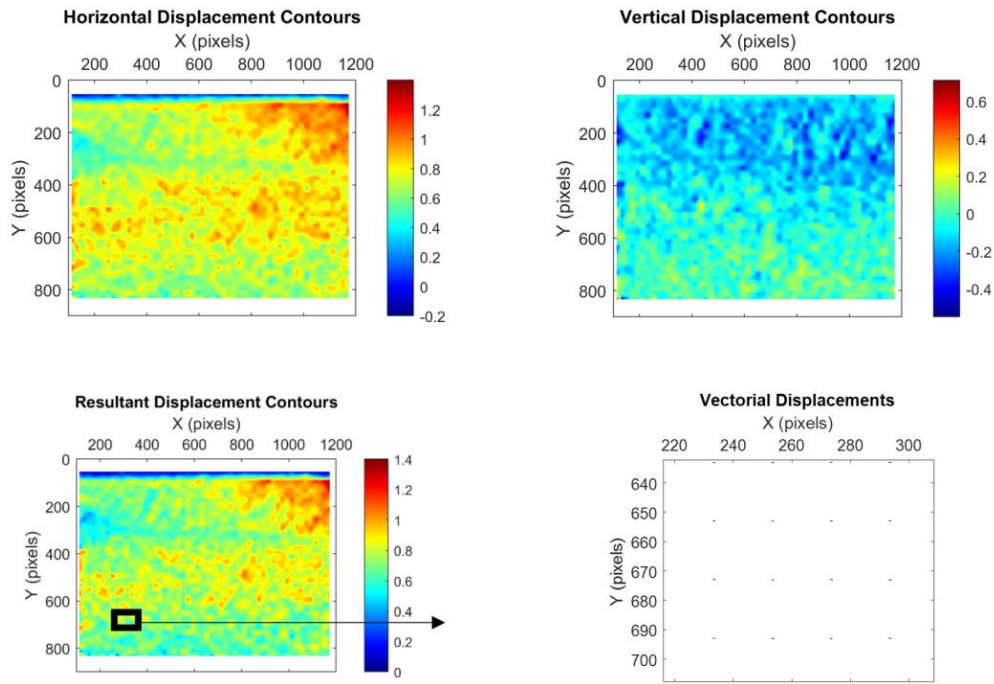

Figure 11 Summary of displacement analysis from GeoPIV a) for experiment 08, b) for experiment 16, and c) for experiment 24

a)
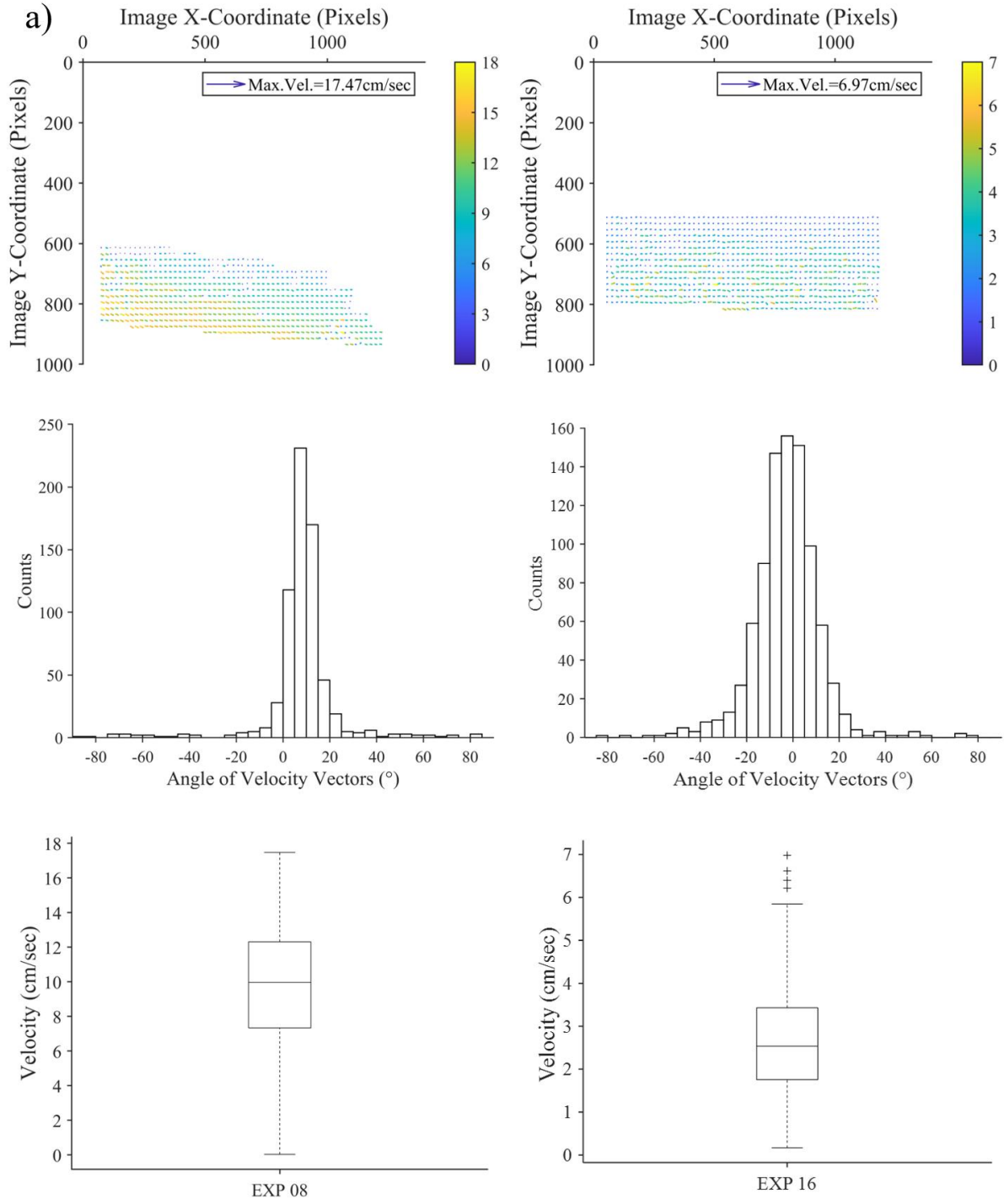

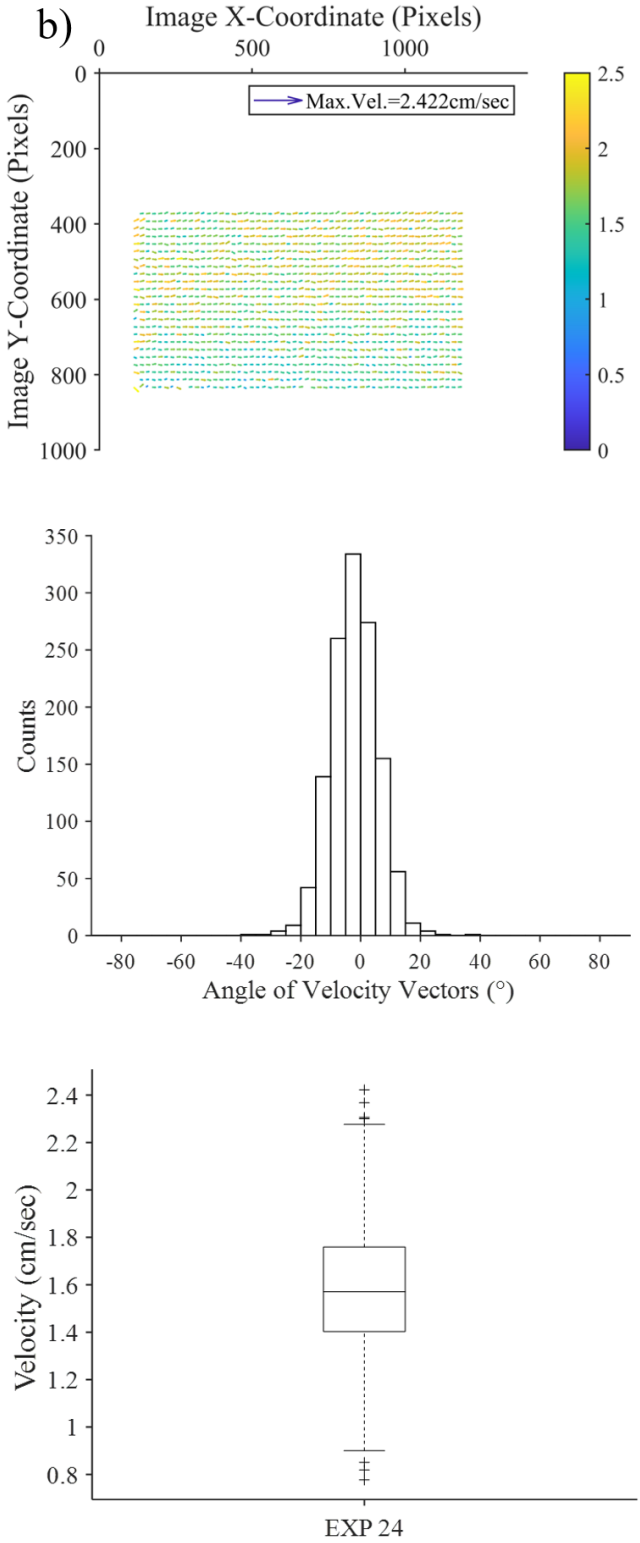

Figure 12a) Summary of post processed velocity analysis for experiments 08 and 16; 12b) for experiment 24

An increase in fracture intersection angles effectively reduces mean and median particle velocities after the intersection region, as shown in Table 5. Generally, increasing intersection angle from 30° to 45° causes a decrease of mean and median particle velocities for about 20% to 30%, which is significant. However, the intersection angle increase causes local swirl effects by observing maximum velocity changes, since maximum velocities could increase up to 60% while it may also decrease to about 40%.

Table 5 Effect of fracture intersection angle on maximum, mean and median velocity

|  | Max Velocity | | Mean Velocity | | Median Velocity | |
| --- | --- | --- | --- | --- | --- | --- |
|  | cm/sec | % Difference | cm/sec | % Difference | cm/sec | % Difference |
| 01 | 15.82 | -9.92% | 7.76 | -33.49% | 7.92 | -49.37% |
| 02 | 14.25 |  | 5.16 |  | 4.01 |  |
| 03 | 9.84 | 0.00% | 5.79 | -25.91% | 6.80 | -20.59% |
| 04 | 9.84 |  | 4.29 |  | 5.40 |  |
| 05 | 11.38 | 36.73% | 8.43 | -25.50% | 8.87 | -20.74% |
| 06 | 15.56 |  | 6.28 |  | 7.03 |  |
| 07 | 14.22 | 9.21% | 9.69 | -6.47% | 10.11 | -5.10% |
| 08 | 15.53 |  | 9.07 |  | 9.59 |  |
| 09 | 3.81 | 12.60% | 1.87 | -23.83% | 1.75 | -28.29% |
| 10 | 4.29 |  | 1.42 |  | 1.26 |  |
| 11 | 5.69 | -38.31% | 2.56 | -37.77% | 2.54 | -40.04% |
| 12 | 3.51 |  | 1.59 |  | 1.52 |  |
| 13 | 3.84 | 60.68% | 1.34 | -14.93% | 1.08 | -23.26% |
| 14 | 6.17 |  | 1.14 |  | 0.83 |  |
| 15 | 5.87 | 18.91% | 1.93 | -19.66% | 1.41 | -27.30% |
| 16 | 6.98 |  | 1.55 |  | 1.03 |  |
| 17 | 3.50 | -28.29% | 1.98 | -23.16% | 1.96 | -26.79% |
| 18 | 2.51 |  | 1.52 |  | 1.44 |  |
| 19 | 3.55 | -10.42% | 2.65 | -28.05% | 2.67 | -29.03% |
| 20 | 3.18 |  | 1.91 |  | 1.90 |  |
| 21 | 2.78 | -16.91% | 1.82 | -24.66% | 1.80 | -24.79% |
| 22 | 2.31 |  | 1.37 |  | 1.35 |  |
| 23 | 3.40 | -24.71% | 2.18 | -28.01% | 2.16 | -29.17% |
| 24 | 2.56 |  | 1.57 |  | 1.53 |  |
| **Average** |  | 0.80% |  | -24.29% |  | -27.04% |

An increase in fluid flow rate will undoubtfully increase maximum, mean, and median particle velocities after the intersection region, as shown in Table 6. Except for some experiments with decreasing effects, particle velocities general increases more than 20% for most of the experiments. In general, while eliminating the effects of proppant concentration, fracture intersection angle, and fluid viscosity, an increase in flow rate (pump rotor frequency from 10Hz to 20Hz, i.e. flow rate increases about 10% to 37%) will averagely cause maximum velocity increase 25%, mean velocity increase 27%, and median velocity increase 31%.

Table 6 Effect of fluid flow rate on maximum, mean and median velocity

|    | Max Velocity | | Mean Velocity | | Median Velocity | |
|----|--------|--------------|--------|--------------|--------|--------------|
|    | cm/sec | % Difference | cm/sec | % Difference | cm/sec | % Difference |
| 01 | 15.82  | -37.8%       | 7.76   | -44.6%       | 7.92   | -31.9%       |
| 03 | 9.84   |              | 4.29   |              | 5.40   |              |
| 05 | 11.38  | 25.0%        | 6.28   | 54.4%        | 7.03   | 43.8%        |
| 07 | 14.22  |              | 9.69   |              | 10.11  |              |
| 02 | 14.25  | -30.9%       | 5.16   | 12.3%        | 4.01   | 69.5%        |
| 04 | 9.84   |              | 5.79   |              | 6.80   |              |
| 06 | 15.56  | -0.2%        | 8.43   | 7.5%         | 8.87   | 8.1%         |
| 08 | 15.53  |              | 9.07   |              | 9.59   |              |
| 09 | 3.81   | 49.3%        | 1.87   | 36.8%        | 1.75   | 44.9%        |
| 11 | 5.69   |              | 2.56   |              | 2.54   |              |
| 13 | 3.84   | 52.9%        | 1.13   | 70.6%        | 1.08   | 31.2%        |
| 15 | 5.87   |              | 1.93   |              | 1.41   |              |
| 10 | 3.51   | 22.2%        | 1.42   | 11.8%        | 1.26   | 21.1%        |
| 12 | 4.29   |              | 1.59   |              | 1.52   |              |
| 14 | 6.17   | 13.1%        | 1.37   | 13.3%        | 0.83   | 24.2%        |
| 16 | 6.98   |              | 1.55   |              | 1.03   |              |
| 17 | 3.50   | 1.43%        | 1.98   | 34.05%       | 1.96   | 36.22%       |
| 19 | 3.55   |              | 2.65   |              | 2.67   |              |
| 21 | 2.78   | 22.30%       | 1.82   | 19.97%       | 1.80   | 20.33%       |
| 23 | 3.40   |              | 2.18   |              | 2.16   |              |
| 18 | 2.51   | 26.69%       | 1.52   | 25.54%       | 1.44   | 32.06%       |
| 20 | 3.18   |              | 1.91   |              | 1.90   |              |
| 22 | 2.31   | 10.82%       | 1.37   | 14.63%       | 1.35   | 13.33%       |

| 24 | 2.56 | 1.57 | 1.53 |
| --- | --- | --- | --- |
| **Average** | 24.9% | 27.4% | 31.3% |

An increase in proppant volumetric concentration has a vague effect in maximum, mean, and median particle velocities as shown in Table 7. An increase in proppant concentration from 10% to 20% will cause both increase and decrease in maximum, mean, and median particle velocities. Most of the cases show decrease (about 10% to 35%) in velocity when proppant volumetric concentration increases, which is in accordance to previous findings. However, there are some extremely increased cases in our experiments, which push the average result towards velocity increase. On average, maximum velocities goes up 15%, mean velocity goes up 8.6%, and median velocity goes up 3%.

Table 7 Effect of proppant volumetric concentration on maximum, mean and median velocity

|  | Max Velocity | | Mean Velocity | | Median Velocity | |
| --- | --- | --- | --- | --- | --- | --- |
|  | cm/sec | % Difference | cm/sec | % Difference | cm/sec | % Difference |
| 1 | 15.82 | -28.1% | 7.76 | -19.1% | 7.92 | -11.3% |
| 5 | 11.38 | | 6.28 | | 7.03 | |
| 3 | 9.84 | 44.5% | 4.29 | 125.8% | 5.40 | 87.3% |
| 7 | 14.22 | | 9.69 | | 10.11 | |
| 2 | 14.25 | 9.2% | 5.16 | 63.5% | 4.01 | 121.2% |
| 6 | 15.56 | | 8.43 | | 8.87 | |
| 4 | 9.84 | 57.8% | 5.79 | 56.6% | 6.80 | 41.1% |
| 8 | 15.53 | | 9.07 | | 9.59 | |
| 9 | 3.81 | 0.8% | 1.87 | -39.4% | 1.75 | -38.6% |
| 13 | 3.84 | | 1.13 | | 1.08 | |
| 11 | 5.69 | 3.2% | 2.56 | -24.4% | 2.54 | -44.4% |
| 15 | 5.87 | | 1.93 | | 1.41 | |
| 10 | 4.29 | 43.8% | 1.42 | -3.7% | 1.26 | -34.3% |
| 14 | 6.17 | | 1.37 | | 0.83 | |
| 12 | 3.51 | 98.9% | 1.59 | -2.4% | 1.52 | -32.6% |
| 16 | 6.98 | | 1.55 | | 1.03 | |
| 17 | 3.50 | -20.57% | 1.98 | -8.10% | 1.96 | -8.42% |
| 21 | 2.78 | | 1.82 | | 1.80 | |
| 19 | 3.55 | -4.23% | 2.65 | -17.75% | 2.67 | -19.10% |

| | | | | | | | |
|---|---|---|---|---|---|---|---|
| 23 | 3.40 | | 2.18 | | 2.16 | | |
| 18 | 2.51 | -7.97% | 1.52 | -9.88% | 1.44 | -5.92% | |
| 22 | 2.31 | | 1.37 | | 1.35 | | |
| 20 | 3.18 | -19.50% | 1.91 | -17.72% | 1.90 | -19.26% | |
| 24 | 2.56 | | 1.57 | | 1.53 | | |
| **Average** | | 14.8% | | 8.6% | | 3.0% | |

An increase in fluid viscosity will effectively decrease particle maximum velocities as shown in Table 8 below. While within more viscous domain, the particle mean and median velocities may also increase a little bit as viscosity increases. Increase fluid viscosity will also cause flow vectors more horizontal.

Table 8 Effect of fluid viscosity on maximum, mean and median velocities

| | Max Velocity | | Mean Velocity | | Median Velocity | | Mode Velocity Angle |
|---|---|---|---|---|---|---|---|
| | cm/sec | Trend | cm/sec | Trend | cm/sec | Trend | degree |
| 01 | 15.82 | keep decreasing | 7.76 | decrease, then increase a little bit | 7.92 | decrease, then increase a little bit | (15,20) |
| 09 | 3.81 | | 1.87 | | 1.75 | | (0,5) |
| 17 | 3.50 | | 1.98 | | 1.96 | | (0,5) |
| 03 | 9.84 | keep decreasing | 4.29 | decrease, then increase a little bit | 5.40 | decrease, then increase a little bit | (15,20)(20,25) |
| 11 | 5.69 | | 2.56 | | 2.54 | | (0,5) |
| 19 | 3.55 | | 2.65 | | 2.67 | | (0,5) |
| 05 | 11.38 | keep decreasing | 6.28 | decrease, then increase | 7.03 | decrease, then increase | (10,15) |
| 13 | 3.84 | | 1.13 | | 1.08 | | (0,5) |
| 21 | 2.78 | | 1.82 | | 1.80 | | (0,5) |
| 07 | 14.22 | keep decreasing | 9.69 | decrease, then increase a little bit | 10.11 | decrease, then increase | (10,15) |
| 15 | 5.87 | | 1.93 | | 1.41 | | (0,5) |
| 23 | 3.40 | | 2.18 | | 2.16 | | (0,5) |
| 02 | 14.25 | keep decreasing | 5.16 | decrease, then increase a little bit | 4.01 | decrease, then increase a little bit | (15,20) |
| 10 | 4.29 | | 1.42 | | 1.26 | | (0,5) |
| 18 | 2.51 | | 1.52 | | 1.44 | | (0,5) |
| 04 | 9.84 | keep decreasing | 5.79 | decrease, then increase | 6.80 | decrease, then increase | (15,20)(20,25) |
| 12 | 3.51 | | 1.59 | | 1.52 | | (0,5) |
| 20 | 3.18 | | 1.91 | | 1.90 | | (0,5) |
| 06 | 15.56 | keep decreasing | 8.43 | decrease, then remain same | 8.87 | decrease, then increase | (10,15) |
| 14 | 6.17 | | 1.37 | | 0.83 | | (0,5) |
| 22 | 2.31 | | 1.37 | | 1.35 | | (0,5) |
| 08 | 15.53 | keep decreasing | 9.07 | decrease, then remain same | 9.59 | decrease, then increase | (5,10) |
| 16 | 6.98 | | 1.55 | | 1.03 | | (0,5) |
| 24 | 2.56 | | 1.57 | | 1.53 | | (0,5) |

Fracture intersection angle has a strong effect on the linear fit slope across different fluid viscosities as shown in Figure 13. For 45° intersection, an increase of 0.001 Pa·s of fluid viscosity will cause an 1108° decrease in the linear fit slope. For 30° intersection, an increase of 0.001 Pa·s of fluid viscosity will cause an 827° decrease in the linear fit slope. Although this difference looks very close, 1108° versus 827°, it is pretty significant compared with the case considering the effect of the fracture intersection angle shown below. Also, an increasing in intersection angle from 30° to 45° will generally cause an increase of 2.1° in slope settlement angle.

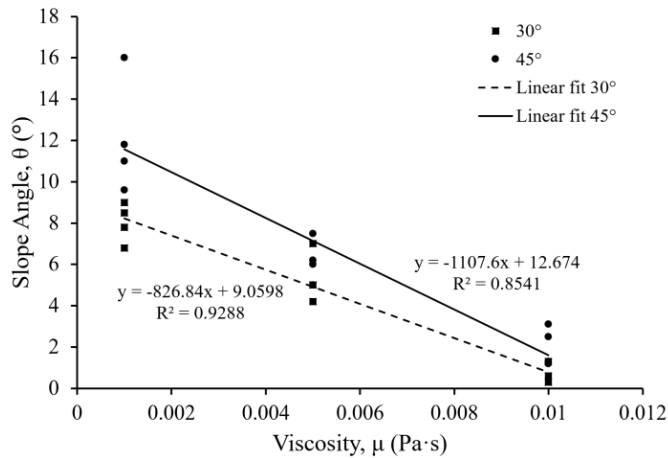

Figure 13 Effect of fracture intersection angle under different fluid viscosities

As expected, proppant particle concentration has a limited effect on settlement slope across different fluid viscosities as shown in Figure 14. For 10 percent particle concentration, an increase of 0.001 Pa·s of fluid viscosity will cause an 844° decrease in the linear fit slope. For 20 percent particle concentration, an increase of 0.001 Pa·s of fluid viscosity will cause an 824° decrease in the linear fit slope. As mentioned previously, this difference, the linear fit slope 844° versus 824°, is much trivial compared with the case considering the effect of fracture intersection angle. Also, an increase in proppant concentration from 10% to 20% will averagely cause slope settlement angle increase 0.2°. Under different combined conditions, the slope settlement angle decreases down to 3.6°, while it can also increase up to 2.5°. There are 6 comparisons with increasing slope trend, 5 comparisons with decreasing trend, and 1 with no decreasing or increasing trend. Therefore, the effect of proppant concentration does not show very strong preference on settlement slope changing trend.

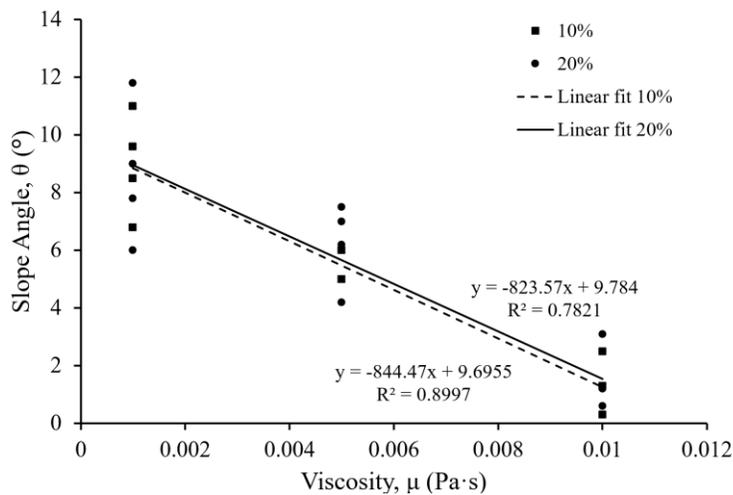

Figure 14 Effect of proppant volumetric concentration

When checking the combined effects of particle concentration and intersection angle, the intersection angle plays a dominant role in slope settlement compared to proppant volumetric concentration as shown in Figures 15a to 15d below. When considering two scenarios, the same intersection angle but difference concentration (Figures 15a and 15b) versus the same concentration but difference intersection angle (Figures 15c and 15d), the latter scenario has bigger differences in slope settlement by observing the inclination differences between fitted lines for each figure.

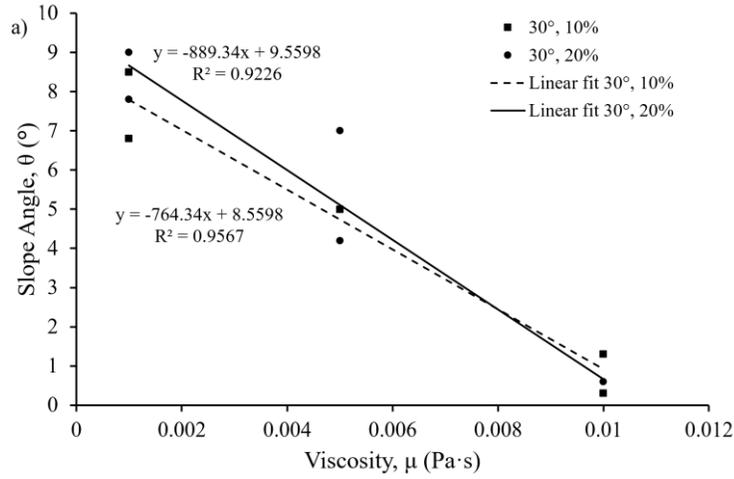

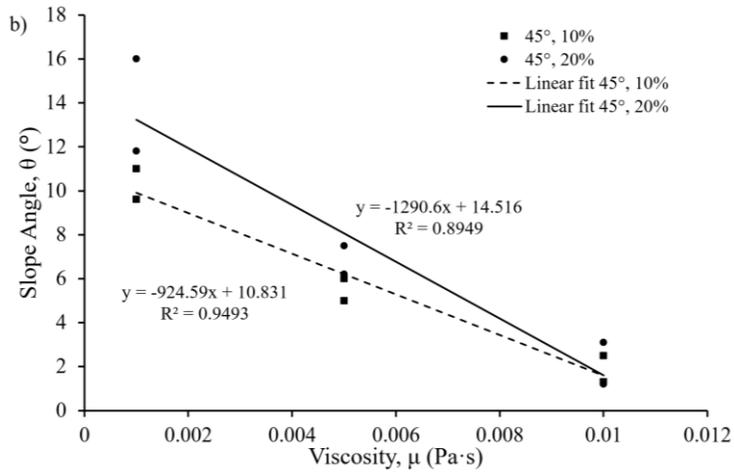

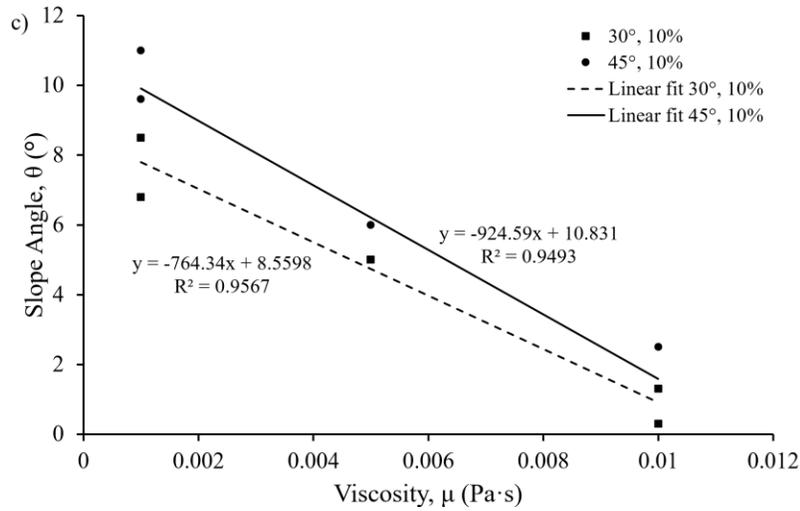

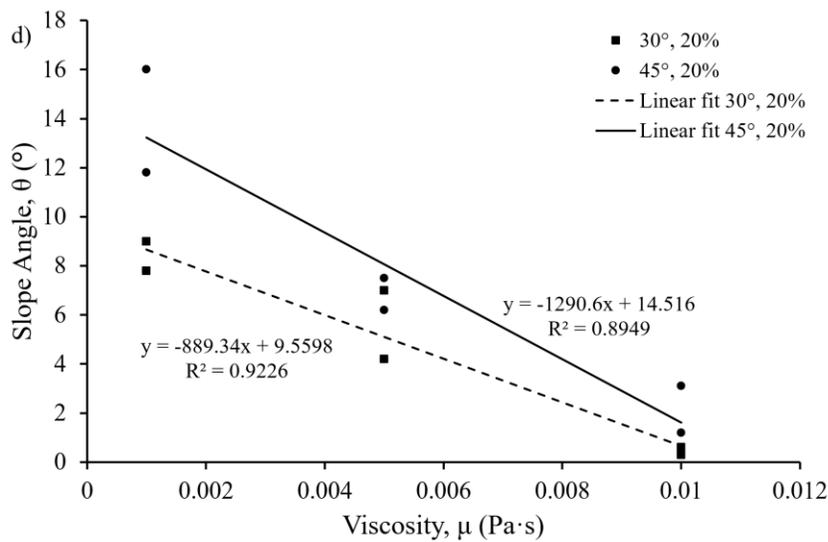

Figure 15a) Combined effect of proppant concentration and fracture intersection at 30 degree fracture intersections but different concentrations; 15b) at 45 degree fracture intersection but different concentrations; 15c) at 10% concentration but different fracture intersection angles; 15d) at 20% concentration but different fracture intersection angles

It is also confirmed that the flow rate has a clear effect on settlement slope across fluid viscosities as shown in Figure 16 below. For flow rate associated with 10 Hz pump rotor frequency, an increase of 0.001 Pa·s of fluid viscosity will cause a 870° decrease in the linear fit slope. For the flow rate associated with 20 Hz pump rotor frequency, an increase of 0.001 Pa·s of fluid viscosity will cause a 1064° decrease in the linear fit slope. Again, comparing this difference, 870° versus 1064° the linear fit slope, with differences under other cases before, it is reasonable to conclude that the flow rate has a clear effect on the linear fit slope. An increase in fluid flow rate will averagely cause an increase of 0.2° in the settlement slope. Under different combined conditions, the slope settlement angle decreases down to 1.7°, while it can also increase up to 4.2°. There are 4 comparisons with increasing slope trend, 7 comparisons with decreasing trend, and 1 with no decreasing or increasing trend. Therefore, the effect of proppant concentration tends to indicate a flattening effect on settlement slope.

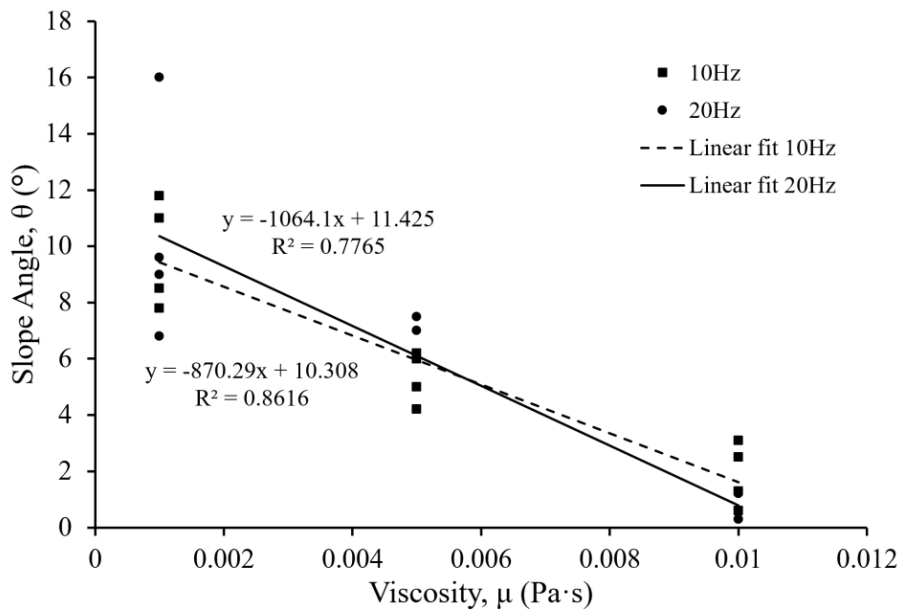

Figure 16 Effect of fluid flow rate

When checking the combined effect of fluid flow rate and intersection angle, it is clear to see that the intersection angle helps to magnify the slope settlement as shown in Figures 17a to 17d below. When considering two scenarios, the same intersection but different flow rate (Figures 17a and 17b) vs. same flow rate but different intersection (Figures 17c and 17d), the latter case causes more significant differences in slope settlement by observing the inclination differences between fitted lines for each figure.

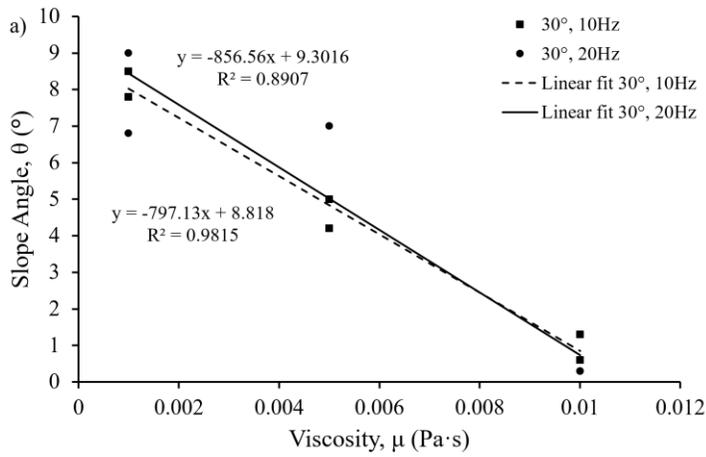

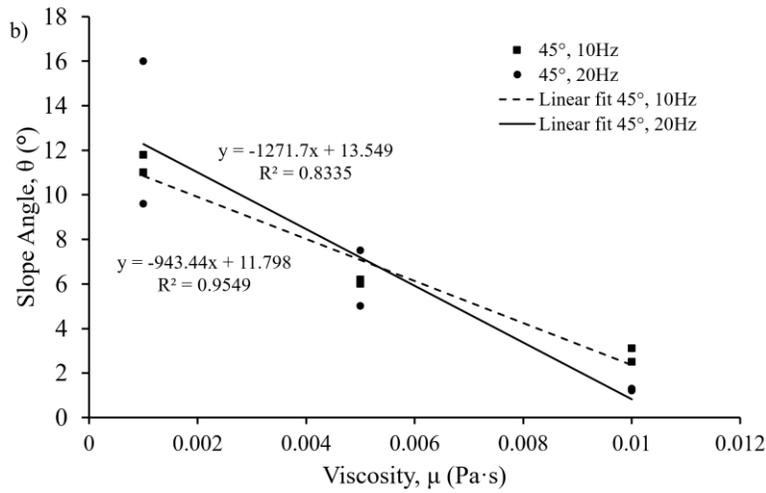

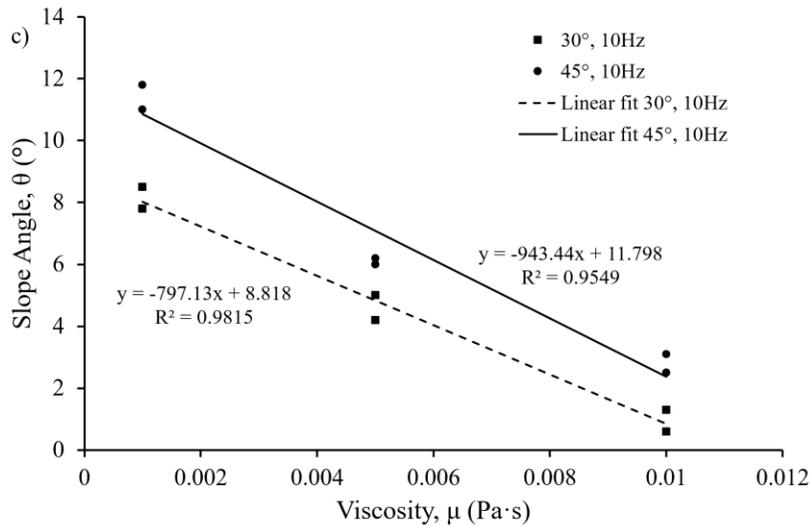

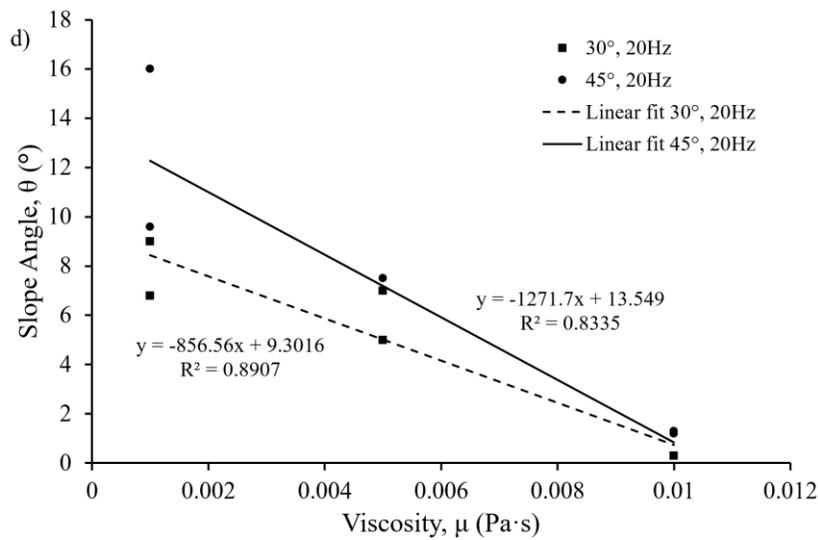

Figure 17 a) Combined effect of fluid flow rate and fracture intersection angle at 30 degree intersection angle but different flow rates; 17b) at 45 degree intersection angle but different flow rates; 17c) at 10Hz pump rotor rate but different fracture intersection angles; 17d) at 20 Hz pump rotor rate but different intersection angles

Slope settlement slope has a logarithmic relationship with particle Reynolds number. The suggested relationships are provided below:

$$Slope(°) = 1.757(°) * \ln(Re) + 3.647(°) \quad \text{for } 30° \text{ intersection} \quad (7a)$$

$$Slope(°) = 2.648(°) * \ln(Re) + 5.865(°) \quad \text{for } 45° \text{ intersection} \quad (7b)$$

The correlation is also shown in Figure 18 below. As particle Reynolds number increases, the settlement slope angle increases accordingly. Also, a higher fracture intersection angle will cause more effect on the slope settlement angle as the particle Reynolds number increases.

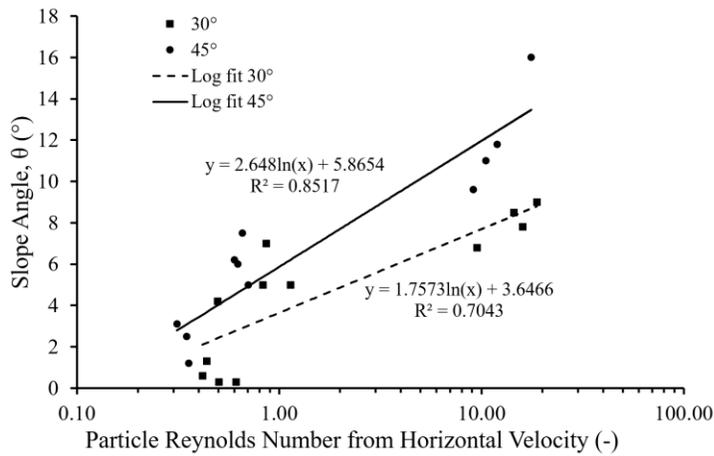

Figure 18 Relationship between slope angle and particle horizontal Reynolds number

Settlement slope also has a logarithmic relationship with particle vertical Stokes number. Suggested relationships are provided below:

$$\text{Slope}(°) = 1.013(°) \cdot \ln(St) + 12.02(°) \quad for\ 30° \ intersection \quad (8a)$$

$$\text{Slope}(°) = 1.865(°) \cdot \ln(St) + 20.34(°) \quad for\ 45° \ intersection \quad (8b)$$

The correlation is also shown in Figure 19 below. As particle Stokes number increases, the settlement slope angle increases accordingly. Also, a higher fracture intersection angle will cause more effect on the slope settlement angle as the particle Stokes number increases.

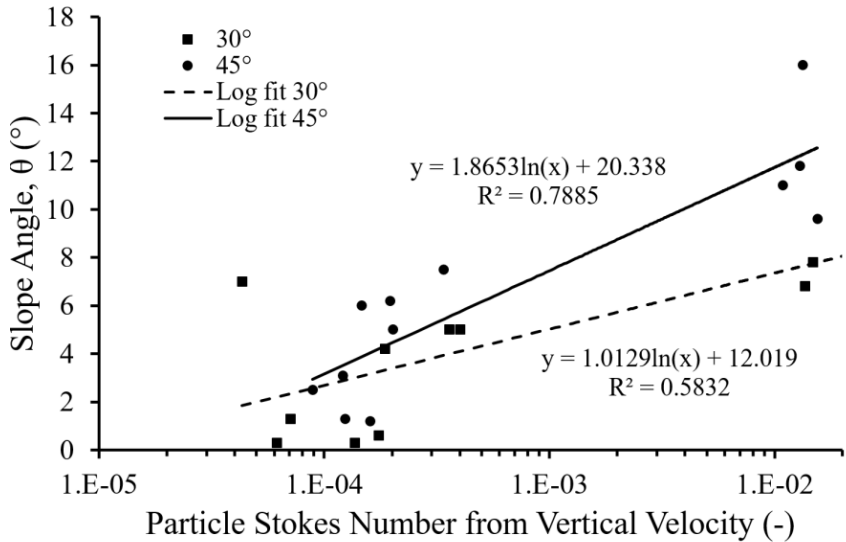

Figure 19 Relationship between slope angle and Stokes number

Supplemental to finding the correlation between the settlement slope angle and an average particle velocity, it is also interesting to examine the correlation between the settlement slope angle and an average particle velocity direction under the effect of fracture intersection angle. Particle velocity direction is compared relative to the horizontal level. A positive number is counted in a clockwise direction. As shown in Figure 20 below, a higher fracture intersection angle will cause a stronger effect on the settlement slope angle as the average particle velocity direction increases. As particle velocity direction increases, the settlement slope angle also increases.

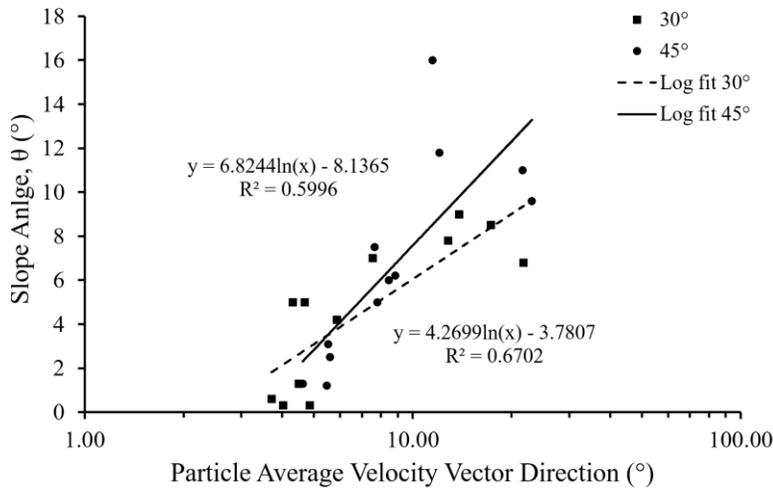

Figure 20 Slope Settlement Angle vs. Average particle velocity Direction

**4.0 Conclusions**

The study uses plexiglas laboratory slot experiments enhanced with advanced image analysis for identifying particle trajectories and quantifying slurry velocities. GeoPIV analysis helps to visualize the particle flow in the fracture and find correlation between settlement slope, particle velocity and velocity direction. GeoPIV enable us to quantify effects of intersection angle, coupled with fluid viscosity, fluid flow rate, and proppant volumetric concentration on proppant velocities. Also, this study aim to find a correlation between particle velocity and slope settlement.

Fracture intersection angle has a strong effect on proppant transport and settlement. Higher fracture intersection angle will generally induce a higher proppant settlement slope angle after the intersection. An increasing in intersection angle from 30° to 45° will cause an increase of 2.1° in slope settlement angle. Higher fracture intersection angle will generally decrease mean and median particle velocity, while creating different effects on maximum velocity. An increase in intersection angle from 30° to 45° causes a decrease of mean and median particle velocities for about 20% to 30%, which is significant. Maximum velocities could increase up to 60% while it may also decrease to about 40%, which suggests local eddies could have occurred.

Changes in proppant concentration have relatively vague effect on slope settlement and particle velocities compared with fluid flow rate. However, intersection angle will help to activate that effect. An increase of proppant concentration from 10% to 20% causes only 0.2° on average increase in settlement slope. However, among all 12 comparisons, a change in proppant concentration has a more balanced preference on increasing or decreasing trend in settlement slope, unlike fluid flow rate. Mean, median and maximum particle velocities recorded in experiments show conflicting and varied values, such as are increase up to about 100% and decrease down to about 40% at higher proppant concentrations. Variations in velocities and the observed swirls suggest that the higher proppant concentration may cause more particle collisions and fluid-particle coupled effects.

Fluid flow rate effects on the slope settlement and particle velocities are more significant than the proppant concentration effects. Higher fluid flow rate results in clear change of the settlement slope. Besides, fracture intersection angle will help to make this effect more significant. An increase in fluid flow rate (pump rotor frequency from 10 Hz to 20 Hz) will averagely cause an increase of 0.2° in slope settlement. However, among all 12 comparisons, a change in fluid flow rate has decreasing-biased effect on settlement slope, unlike proppant concentration. An increase in fluid flow rate increases maximum, median, and mean particle velocities for 25%, 31%, and 27% respectively.

Fluid viscosity effect flattens the proppant settlement slope after intersected fracture and helps to transport proppant further to the exit. However, fracture intersection angle increase counter-balances this flattening effect. Particle velocities significantly decrease under lower viscosity range, while they will remain similar or increase slightly under higher viscosity range.

**CRediT Author Statement**

**Wenpei Ma:** Conceptualization, Methodology, Investigation, Formal Analysis, Writing – Original Draft;

**Justin Perng:** Investigation, Formal Analysis;

**Ingrid Tomac:** Conceptualization, Methodology, Writing – Review & Editing, Supervision, Project Administration, Funding Acquisition.

**Acknowledgments**

This work was supported by the U.S. National Science Foundation, Division of Civil, Mechanical and Manufacturing Innovation [grant number NSF CMMI 1563614] and University of California, San Diego.